\documentclass[journal, romanappendices, final]{IEEEtran}
\usepackage{latexsym,amsfonts,amssymb,amsthm, setspace, verbatim,cite,mathrsfs,graphicx,bm,stfloats, hyperref, tikz,xcolor, bm, algorithmic, algorithm,  enumerate}
\usepackage[cmex10]{amsmath}
\newcommand{\un}{\underline}
\newcommand{\be}{\begin{equation}}
\newcommand{\ee}{\end{equation}}
\newcommand{\ben}{\begin{equation*}}
\newcommand{\een}{\end{equation*}}
\newcommand{\mc}{\mathcal}
\newcommand{\mbf}{\mathbf}
\newcommand{\e}{\epsilon}
\newcommand{\abs}[1]{\lvert#1\rvert}

\newcommand{\expec}{\mathbb{E}}
\newtheorem{lem}{Lemma}

\newtheorem{defi}{Definition}
\newtheorem{thm}{Theorem}
\newtheorem{fact}{Fact}

\newtheorem{prop}{Proposition}

\begin{document}
\title{Low-Complexity Interactive Algorithms for  Synchronization from Deletions, Insertions, and Substitutions}
\author{Ramji Venkataramanan,~\IEEEmembership{Member,~IEEE,}
Vasuki Narasimha Swamy,~\IEEEmembership{Student Member,~IEEE,}
and \\ Kannan Ramchandran,~\IEEEmembership{Fellow,~IEEE \vspace{-10pt}}
\thanks{The material in this paper was presented in part at the 2010 and 2013 Allerton Conference on Communication, Control, and Computing.}%
\thanks{R.~Venkataramanan is with the Department of Engineering, University of Cambridge, Cambridge CB2 1PZ, UK (e-mail: ramji.v@eng.cam.ac.uk).}%
\thanks{V. N.~Swamy and K.~Ramchandran are with Department of Electrical Engineering and Computer Sciences, University of California, Berkeley, CA
94720 USA (e-mail: \{vasuki,kannanr\}@eecs.berkeley.edu).}
%
%
}
\maketitle

\begin{abstract}
Consider two remote nodes having binary sequences $X$ and $Y$, respectively. $Y$  is an \emph{edited} version of  ${X}$, where the editing involves random  deletions, insertions, and substitutions, possibly in  bursts. The goal is for the node with $Y$  to reconstruct $X$ with minimal exchange of information over a noiseless link. The communication is measured in terms of both the total number of bits exchanged and the number of interactive rounds of communication. 

This paper focuses on the setting where the number of edits is $o(\tfrac{n}{\log n})$, where $n$ is the length of $X$. We first consider the case where the edits are a mixture of insertions and deletions (indels), and propose an interactive synchronization algorithm with near-optimal communication rate and average computational complexity of $O(n)$ arithmetic operations. The algorithm uses interaction to efficiently split the source sequence into substrings containing exactly one deletion or insertion.  Each of these substrings is then synchronized using an optimal one-way synchronization code based on the single-deletion correcting channel codes of Varshamov and Tenengolts (VT codes).

We then build on this synchronization algorithm in three different ways. First, it is modified to work with a single round of interaction. The reduction in the number of rounds comes at the expense of higher communication,  which is quantified.  Next, we present an extension to  the practically important case where the  insertions and deletions may occur in (potentially large) bursts. Finally,  we show how to synchronize the sources to within a target Hamming distance. This feature can be used to differentiate between substitution and indel edits. In addition to theoretical performance bounds,  we provide several validating simulation results for the proposed algorithms.
\end{abstract}

\begin{IEEEkeywords}
File Synchronization, Two-way interaction,   Deletion edits, Insertion edits, Edit channels, Varshamov-Tenengolts codes
\end{IEEEkeywords}

\section{Introduction}
\label{sec:intro}
\IEEEPARstart{C}onsider two remote nodes, say Alice and Bob, having binary sequences $X$ and $Y$, respectively. $Y$ is an edited version of $X$, where the edits may consists of deletions, insertions, and substitutions of bits.  Neither party knows what has been edited nor the locations of the edits. The goal is for Bob to reconstruct Alice's sequence  with minimal communication between the two. This  problem  of efficient synchronization arises in  practical applications such as file backup (e.g., Dropbox), online file editing, and file sharing. rsync~\cite{rsync} is a UNIX utility that can be used to synchronize two remote files or directories. It uses hashing to determine the parts where the two files match, and then transmits the parts that are different. Various forms of the file synchronization problem have been studied in the literature, see e.g., \cite{evf98,CormodePSV00,OrlitskyV01,Trachten06,DolBit13,MaKTse12}.

\begin{figure*}[t]
\centering
\includegraphics[width=6.5in]{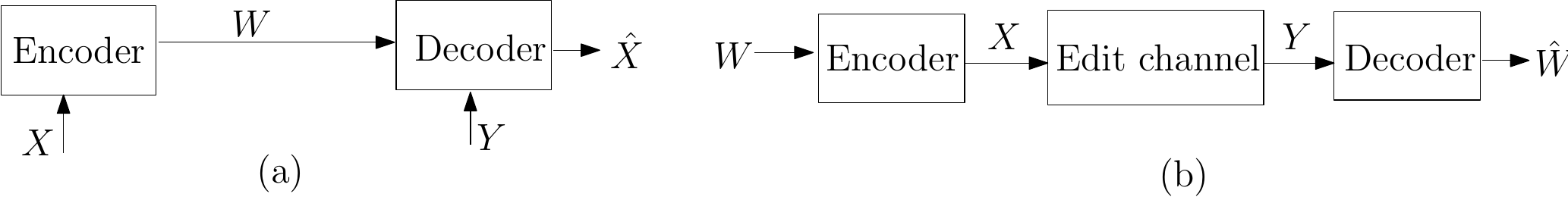}
\caption{\small{ (a) Synchronization: reconstruct $X$ at the decoder using the message $W$ and the edited version $Y$ as side-information.
 (b) Channel coding: transmit message $W$ through a channel that takes input $X$ and outputs edited version $Y$.
 \vspace{-5pt} }}
\label{fig:dual_problems}
\end{figure*}

In this paper,  we propose synchronization algorithms and analyze their performance for the setting where the total number of edits is small compared to the file size. In particular, we focus on the case where the number of edits $t = o(\tfrac{n}{\log n})$, where $n$ is the length of Alice's sequence $X$.
From here on, we will refer to Alice and Bob as the encoder and decoder, respectively (Fig. \ref{fig:dual_problems}(a)). We assume that the lengths of $X$ and $Y$ are known to both the encoder and decoder at the outset.

 A natural first question is: what is the minimum communication required for synchronization? A simple lower bound on the  communication required to synchronize from insertion and deletion edits can be obtained by assuming that the encoder knows the locations of the $t$ edits in $X$. Then, the minimum number of bits needed to convey the positions of the edits to the decoder  is approximately $t \log n$ ($\approx \log n$ bits to indicate each position). This is discussed  in more detail in Section \ref{sec:limits}.

When $X$ and $Y$ differ by exactly one deletion or insertion, there is a simple one-way, zero error algorithm to synchronize  $Y$ to $X$. This algorithm, based on a family of single-deletion correcting codes introduced by Varshamov and Tenengolts~\cite{VT65}, requires $\log(n+1)$ bits to be transmitted from the encoder to the decoder, which is very close to the lower bound of $\log n$. However, when $X$ and $Y$ differ by multiple deletions and insertions, there is no known one-way synchronization algorithm that is computationally feasible and transmits significantly fewer than $n$ bits.

In this work, we insist on realizable (practical) synchronization algorithms, and relax the requirement of zero error---we will only require that the probability of synchronization error goes to zero polynomially in the problem size $n$. Specifically, we develop a fast synchronization algorithm   by allowing a small amount of \textit{interaction} between the encoder and the decoder.  When the number of edits $t = o(\frac{n}{\log n})$, the total number of bits transmitted by this algorithm is within a constant factor of the communciation lower bound $t \log n$, where the constant controls the polynomial  rate of decay of the probability of synchronization error. To highlight the main ideas and keep the exposition simple, we focus on the case where $X$ and $Y$ are binary sequences. All the algorithms can be extended in a straightforward manner to larger discrete alphabets; this is briefly discussed in Section  \ref{sec:concl}.

We lay down some notation before proceeding. Upper-case letters are used to denote random variables and random vectors, and lower-case letters for their realizations. $\log$ denotes the logarithm with base $2$, and $\ln$ is the natural logarithm.  The length of $X$ is denoted by $n$, and the number of edits is denoted by $t$. We use $N_{1 \to 2}$ to denote the number of bits sent from the encoder to the decoder, and $N_{2 \to 1}$ to denote the number of bits sent by the decoder to the encoder.

Following standard notation, $f(n)=o(g(n))$ means $\lim_{n \to \infty} f(n)/g(n) =0$; $f(n)=O(g(n))$  means  $f$ is asymptotically bounded above by $\kappa g(n)$ for some constant $\kappa >0$, and $f(n)=\Theta(g(n))$ means $f(n)/g(n)$ asymptotically lies in an interval $[\kappa_1,\kappa_2]$ for some constants $\kappa_1,\kappa_2>0$.

\subsection{Contributions of the paper}
 The main contribution is a bi-directional algorithm to synchronize from an arbitrary combination of insertions and deletions (referred to hereafter as \emph{indels}). 
 For the case where $X$  is a uniform random binary string and $Y$ is  obtained from $X$ via  $t$  deletions and insertions whose locations are uniformly  random, the expected number of bits transmitted by the algorithm from the encoder  to the decoder  is close to $4 c t\log n$,  where $c > 1.5$ is a  user-defined constant that controls the trade-off between the communication required and probability of synchronization error.  The expected number of bits in the reverse direction is approximately $10 t$.  Therefore the total number of bits exchanged between encoder and decoder is within a constant factor of the lower bound $t \log n$. The probability of synchronization error goes to zero as $\frac{t\log n}{n^c}$. The synchronization algorithm has  average computational complexity of   $O(n)$ arithmetic operations, with $O(\log n)$ bits of memory.

We then present three extensions:

\begin{enumerate}
\item \emph{Limited number of rounds}:   The number of rounds in the bi-directional synchronization algorithm  of  is  of the order of 
$\log t$, where $t$ is the number of indel edits. In practical applications where the sources may be connected by a high-latency link, having a large number of  interactive rounds is not feasible --- rsync, for example, uses only one round of interaction. In Section \ref{sec:limited_rounds}, we modify the algorithm to work with only one complete round of interaction, and analyze the communication required when $Y$ is generated from $X$ via indel edits at uniformly random locations. Simulation results show that the single-round algorithm  is very fast and requires significantly less communication than rsync.

\item \emph{Bursty Edits}:  In practice, edits in files often occur  in (possibly large) bursts.   For reasons discussed in Section \ref{sec:bursts}, the  performance of the original algorithm is suboptimal for bursty indel edits. To address this, we describe a technique to efficiently synchronize from a single large burst deletion or insertion. We then use this technique in the original algorithm to synchronize efficiently when the edits are a combination of isolated deletions and insertions and bursts of varying length.

\item \emph{Substitution Edits}: In Section \ref{sec:sub_error}, we show how the interactive synchronization algorithm can handle substitution edits in addition to indels. This is done by using a Hamming-distance estimator as a hash in the original synchronization algorithm. This lets us synchronize $Y$ to within a target Hamming distance of $X$. The remaining substitution errors can then be corrected using standard methods based on syndromes of an appropriate linear error-correcting code.
\end{enumerate}

For general files and edit models, several authors \cite{evf98,CormodePSV00,OrlitskyV01,Trachten06} have proposed file synchronization protocols with communication complexity  bounds of $O(t \log^2 n)$ or higher, where $t$ is the edit distance between $X$ and $Y$. In contrast, the main focus of this paper is to design practically realizable synchronization algorithms for binary files with indel edits,  with sharp performance guarantees when the number of edits $t=o(\tfrac{n}{\log n})$. 
 Our main theoretical contributions are for the interactive synchronization algorithm and its  single-round adaptation, under  the assumption that the binary strings and the locations of the edits are uniformly random.  For the case with bursty edits, we provide a theoretical analysis  for the special case of a single burst of deletions or insertions. For the practically important case of multiple edits (including bursts of different lengths), the performance is demonstrated via several simulation results. Likewise, the effectiveness of the Hamming-distance estimator when the edits include substitutions is  illustrated via simulations.

 While the simulations are performed on randomly generated binary strings, this is the first step towards the larger goal of designing a practical rate-efficient  synchronization tool for applications such as video. Indeed, a key motivation for this work was to explore the use of interaction and coding to enhance VSYNC \cite{ZhangYR12}, a recent algorithm for video synchronization.

A preliminary version of the main synchronization algorithm proposed in this paper was presented in the 2010 Allerton conference. The algorithm has subsequently been used as a building block in  other problems including: a) computationally feasible synchronization in the regime where the number of edits grows linearly in $n$ \cite{YazdiDol14,DolBit13}, and b) synchronizing rankings between two remote terminals \cite{sumilenkovic14}.
A brief description of these papers is given at the end of the subsection below.

\subsection{Related work}

When $X$ and $Y$ differ by just substitution edits, the synchronization problem is well-understood: an elegant and rate-optimal one-way synchronization code can be  obtained using cosets of a linear error-correcting code, see e.g. \cite{Wyner74, Orlitsky93, PradhanR03, XiongDSC04}. For general edits, Orlitsky \cite{Orlitsky93} obtained several interesting bounds on the number of bits needed when the number of rounds of communication is constrained. In particular, a three-message algorithm  that synchronizes from $t$ indel edits with a near-optimal number of bits was proposed in \cite{Orlitsky93}.  This algorithm is not computationally feasible, but  for the special case where $X$ and  $Y$ differ by one edit, \cite{Orlitsky93} described a computationally computationally efficient one-way algorithm based on Varshamov-Tenengolts (VT) codes. This algorithm is reviewed in Section  \ref{sec:single_ins_del}.

Evfimievski \cite{evf98} and Cormode et al. \cite{CormodePSV00} proposed different $\e$-error synchronization protocols for which the number of transmitted bits is  
$t \cdot \text{poly}( \log n, \log \e^{-1})$ where  $t$ is the edit distance between $X$ and $Y$,  $\e$ is the probability of synchronization error, and 
$\text{poly}(\log n, \log \e^{-1})$ denotes a polynomial in $\log n$ and $ \log \e^{-1}$. These protocols have computational complexity that is polynomial in $n$. Subsequently, Orlitsky and Viswanathan developed a practical $\e$-error protocol \cite{OrlitskyV01} which communicates  $O(t \log n ( \log n + \log \e^{-1}))$ bits and has $O(n \log n)$ computational complexity.

Agarwal et al. \cite{Trachten06} designed a synchronization algorithm using the approach of \emph{set reconciliation}: the idea is to divide each of  $X$ and $Y$ into overlapping substrings and  to first convey the substrings of $X$ which which differ from $Y$; reconstructing $X$ at the decoder then involves finding a unique Eulerian cycle in a de Bruijn graph. A computationally feasible algorithm for the second step that guarantees reconstruction with high probability is described in \cite{KTracht13}. The communication is $O(t \log^2 n)$ bits when $X$ and $Y$ are random i.i.d strings differing by $t$ edits.

 In \cite{YazdiDol14}, Yazdi and Dolecek consider the problem of synchronization when $Y$ is obtained from $X$ by a process that deletes each bit independently with probability $\beta$.  ($\beta$ is a small constant, so the number of deletions is $\Theta(n)$.) The synchronization algorithm described in Section \ref{sec:insdel} for $o(\tfrac{n}{\log n})$ edits is a key ingredient of the synchronization protocol proposed in \cite{YazdiDol14}.  This protocol transmits a total of $O(n\beta \log \tfrac{1}{\beta})$ bits and the probability of synchronization failure falls exponentially in $n$; the computational complexity is $O(n^4\beta^6)$.  The synchronization protocol in \cite{YazdiDol14} is generalized in \cite{DolBit13} to deal with the case where the  alphabet is non-binary and the edits include both deletions and insertions.  The performance of this protocol is evaluated in \cite{BitouzeSYD13}, and significant gains over rysnc are reported for the setting where $X$ undergoes a constant rate of i.i.d. edits to generate $Y$. In \cite{sumilenkovic14}, an interactive algorithm based on VT codes is proposed for the problem of synchronizing rankings between two remote terminals.
 
The paper is organised as follows. In Section \ref{sec:limits}, we derive a simple lower bound on the minimum communication required to synchronize from $t$ indel edits. In Section \ref{sec:single_ins_del}, we describe how to optimally synchronize from one deletion or insertion. This technique is a key ingredient of the interactive algorithm to synchronize from multiple indel edits, which is described in Section \ref{sec:insdel}. In Sections \ref{sec:limited_rounds}, \ref{sec:bursts}, and \ref{sec:sub_error}, we extend the main synchronization algorithm to work with a single round of interaction, bursty edits, and substitution edits, respectively. Section \ref{sec:proofs} contains the proofs of the main results. Section \ref{sec:concl} concludes the paper.

\section{Fundamental Limits} \label{sec:limits}
The goal in this section is to obtain a lower bound on the minimum number of bits required for synchronization when $X$ and $Y$ differ by $t$ indel edits, where $t = o(n)$. Though similar bounds can be found in \cite[Section $5$]{Orlitsky93},  for completeness we present a bound tailored to the synchronization framework considered here. We begin with the following fact.
\begin{fact}
(a)  Let $Q_t(y)$ denote the number of different sequences that can  be obtained by inserting $t$ bits in length-$m$ sequence $y$. Then,
\be
\begin{split}
Q_t(y) & =  \sum_{l =0}^t {m +t \choose l}. 
\end{split}
\label{eq:qx_bound}
\ee
\label{fact:levPQ}

(b)  For any binary sequence $y$, let $P_t(y)$ denote the number of different sequences that can  be obtained by deleting $t$ bits from $y$. Then,
\be P_t(y) \geq {R(y) - t +1 \choose t}, \label{eq:px_bound} \ee
where $R(y)$ denotes the numbers of runs in $y$.\footnote{The runs of a binary sequence are its alternating blocks of contiguous zeros and ones.}
\end{fact}

Part ($a$) is Lemma $4$ in \cite{Orlitsky93}. Part ($b$) was proved in \cite{Lev65} and can be obtained as follows.
Consider deleting $t$ bits from $Y$ by choosing $t$ non-adjacent runs of $Y$ and deleting one bit  from each of them. Each choice of $t$ non-adjacent runs yields a unique  length-$n$ sequence $X$. The number of ways of choosing $t$ non-adjacent runs from $R(Y)$ runs is given by the right side of \eqref{eq:px_bound}.
Note that the number of sequences that can be obtained by deleting $t$ bits from $Y$ depends on the number of runs in $Y$---for example, deleting any $t$ bits from the all-zero sequence yields the same sequence. The following lemma shows that a large fraction of sequences in $\{0,1\}^m$ have close to $\tfrac{m}{2}$ runs; this will help us obtain a lower bound on the number of bits needed to synchronize ``typical" $Y$-sequences from $t$ insertions.

\begin{lem}
For any $\e \in (0, 1)$,  there are at least $(1-\e)2^{m}$ length-$m$ binary sequences with at least $\tfrac{m}{2} (1- \Delta_{m,\e})$ runs each, where
$\Delta_{m,\e} =  \sqrt{\frac{2}{m-1} \ln  \frac{1}{\e}}$.
\label{lem:run_lem}
\end{lem}
\begin{IEEEproof}
In Appendix \ref{app:run_lem_proof}.
\end{IEEEproof}

The following proposition establishes a lower bound on the number of bits needed for synchronization, and provides a benchmark to compare the performance of the algorithms proposed in this paper.

\begin{prop}
Let $m$ denote the length of the decoder's sequence $Y$. Then any synchronization algorithm that is successful for all length-$n$ 
$X$-sequences compatible with $Y$ satisfies the following. 

(a)  For $m=n-t$, the number of bits the encoder must transmit to synchronize $Y$ to  $X$, denoted by $N_d(n,t)$,  satisfies
\be
\liminf_{n \to \infty} \  \frac{N_d(n,t)}{t \log \left(\frac{n}{t} \right)} \geq1.
\label{eq:ndnt}
\ee
for  $t=o(n)$.

(b) For any $\e \in (0, 1)$,  let $\mc{A}_{\e, m} \subset \{0,1\}^{m}$ be the set of sequences of length $m$ that have at least $\tfrac{m}{2}(1 - \Delta_{m,\e} )$ runs, where $\Delta_{m,\e}=  \sqrt{\tfrac{2}{m-1} \ln \tfrac{1}{\e}}$. Then $\mc{A}_{\e,m}$ has at least $(1-\e)2^m$ sequences. For $m=n+t$,  the number of bits  the encoder must transmit to  synchronize  $Y \in \mc{A}_{\e,m}$ to $X$, denoted by $N_i(n,t)$,  satisfies
 \be
\liminf_{n \to \infty} \   \frac{N_i(n,t)}{t \log \left(\frac{n}{t}\right) } \geq 1.
 \label{eq:nint}
 \ee
 for $t=o(n)$.
\label{prop:prop_lb}
\end{prop}
\emph{{Remark}}: Proposition \ref{prop:prop_lb} assumes that $Y$ is available a priori at the encoder, so the lower bound on the communication required applies to both interactive and non-interactive synchronization algorithms.

\begin{IEEEproof}
(\emph{a}):  Fact \ref{fact:levPQ} (a) implies that the number of possible  length-$n$ $X$ sequences  consistent with any length $(n-t)$ sequence $Y$ is greater than ${n-t +t \choose t}$. Thus we need the encoder to send at least $\log {n \choose t}$ bits, {even} with perfect knowledge of $Y$. To obtain \eqref{eq:ndnt}, we  bound  $ {n \choose t}$ from below using the following bounds  (Stirling's approximation) for the factorial:
\be
\sqrt{2\pi} \, n^{n+\frac{1}{2}} e^{-n} \leq n! \leq  e \, n^{n+\frac{1}{2}} e^{-n}, \quad n \in \mathbb{N}.
\label{eq:stirling}
\ee

(\emph{b}): For any  $\e>0$,  Lemma \ref{lem:run_lem} shows that there are at least $(1-\e)2^m$ length $m$ sequences with at least $\tfrac{m}{2}(1 - \Delta_{m,\e} )$ runs.
 Part ($b$) of Fact \ref{fact:levPQ}  gives a lower bound on the number of possible $X$ sequences consistent with $Y$.   Lemma \ref{lem:run_lem} and Fact \ref{fact:levPQ} together imply that to synchronize any $Y \in \mc{A}_{\e, m}$ where $m=n+t$, the encoder needs to send at least
 \[ \log {\tfrac{m}{2}(1 - \Delta_{m,\e} ) -t +1  \choose t} \text{ bits}, \] even with perfect knowledge of $Y$. Using \eqref{eq:stirling} to bound the factorials and simplifying yields \eqref{eq:nint}.
\end{IEEEproof}

\section{Synchronizing from One Deletion/Insertion} \label{sec:single_ins_del}
In this section, we describe how to optimally synchronize from a single deletion or insertion. The one-way synchronization algorithm  for a single deletion is based on the family of single-deletion correcting channel codes introduced by Varshamov and Tenengolts~\cite{VT65} (henceforth abbreviated to VT codes). 

\begin{defi}\label{def:vt_codes}
For block length $n$, and integer $a \in \{ 0, \ldots, n \}$, the  VT code $VT_a(n)$ consists of all binary vectors $X=(x_1,\ldots,x_n)$ satisfying
\be
\sum_{i=1}^n ix_i \equiv a\;\text{mod}~(n+1).
\ee
\end{defi}
For example, the code $VT_0(4)$  with block length $n=4$ is
\be
\begin{split}
VT_0(4) &= \{(x_1,x_2,x_3,x_4) :  \sum_{i=1}^4 ix_i\;\text{mod}~5 =0 \} \\
& = \{0000, 1001, 0110, 1111 \}.
\end{split}
\ee
 For any $a \in \{0,\ldots, n\}$, the code $VT_a(n)$ can be used to communicate reliably over an edit channel (Fig. \ref{fig:dual_problems}(b)) that introduces at most one deletion in a block of length $n$. Levenshtein proposed a simple decoding algorithm~\cite{Lev65,Sloane00} for a VT code, which we reproduce below. Assume that the channel code $VT_a(n)$ is used.
\begin{itemize}
\item Suppose that a codeword $X \in VT_a(n)$ is transmitted, the channel deletes the bit in position $p$, and $Y$ is received.
Let there be $L_0$ $0$'s and $L_1$ $1$'s to the left of the deleted bit, and $R_0$ $0$'s and $R_1$ $1$'s to the right of the deleted bit (with $p = 1 + L_0 + L_1$).

\item The channel decoder computes the weight  of $Y$ given by $wt(Y) = L_1 + R_1$, and the new checksum
$\sum_{i} i y_i$. If the deleted bit is $0$, the new checksum is smaller than the checksum of $X$ by an amount $R_1$. If the deleted bit is $1$, the new checksum is smaller by an amount $p+R_1 = 1+L_0+L_1+R_1 = 1+wt(Y)+L_0$.

Define the \emph{deficiency} $D(Y)$ of the new checksum as  the amount by which it is smaller than the next larger integer of the form $k(n+1) + a$, for some integer $k$. Thus,  if a $0$ was deleted the deficiency  $D(Y)=R_1$, which is less than $wt(Y)$; if a $1$ was deleted $D(Y) = 1+wt(Y)+L_0$, which is greater than $wt(Y)$.

\item If the deficiency $D(Y)$ is less than or equal to $wt(Y)$ the decoder determines that a $0$ was deleted, and restores it just to the left of the rightmost $R_1$ $1$'s. Otherwise a $1$ was deleted and the decoder  restores it just to the right of the leftmost $L_0$ $0$'s.
\end{itemize}

As an example, assume that the code $VT_0(4)$  is used and $X = (1,0,0,1) \in VT_0(4)$ is transmitted over the channel.
\begin{enumerate}
\item If the second bit in $X$ is deleted and $Y=(1,0,1)$, then the new checksum is $4$, and the deficiency  $D=5-4=1\, < wt(Y)=2$. The decoder inserts a zero just to the left of $D=1$ ones from the right to get $(1,0,0,1)$.
\item If the fourth bit in $X$ is deleted and $Y=(1,0,0)$, then the new checksum is $1$, and the deficiency $D=5-1=4\, > wt(Y)=1$. The decoder inserts a one after $D-wt(Y)-1 = 2$ zeros from the left to get $(1,0,0,1)$.
\end{enumerate}
Note that in the first case, the zero is restored in the third position though the original deleted bit may have been the one in the second position. The VT code implicitly exploits the fact that a deleted bit can be restored at any position within the correct run. The decoding algorithm  always restores a deleted zero at the end of the run it belongs to, while a deleted one is always restored at the beginning of the run.

Several interesting properties of VT codes are discussed in \cite{Sloane00}. For example,  it is known that $VT_0(n)$ is the largest single-deletion correcting code for block lengths $n \leq 9$, i.e., they are rate-optimal for $n \leq 9$ . Further, for each $n$, $VT_0(n)$ has size  at least $\frac{2^n}{n+1}$, which is asymptotically optimal \cite[Corollary 2.3, Theorem 2.5]{Sloane00}.

\subsection{One-way Synchronization using VT Syndromes}
As observed in \cite{Orlitsky93}, VT codes can be used to synchronize from a single deletion. In this setting (Fig. 1(a)), the length-$n$ sequence $X$ is available at the encoder, while the decoder has $Y$, obtained by deleting one bit  from $X$. To synchronize, the encoder sends the checksum of its sequence $X$ modulo $(n+1)$. The decoder receives this value, say $a$, and decodes its sequence $Y$ to a codeword in $VT_a(n)$.  This codeword is equal to $X$ since $VT_a(n)$ is a single-deletion correcting channel code.

 Since $a\in\{0,\ldots,n\}$,  the encoder needs to transmit $\log (n+1)$ bits. This is asymptotically optimal as the lower bound of Proposition \ref{prop:prop_lb} for $t=1$ is $\log n$. We have achieved synchronization  by using the fact that the $\{0,1\}^n$ space is partitioned by the (non-linear) codes $VT_a(n),\, 0 \leq a \leq n$. This is similar to using
cosets of a linear code to perform Slepian-Wolf coding \cite{Wyner74,PradhanR03}. Hence we shall refer to $\sum_{i}ix_i \text{ mod } (n+1)$ as the \emph{VT syndrome} of $X$.

If $Y$ was obtained from $X$ by a single {insertion}, one can use a similar algorithm to synchronize $Y$ to $X$. The only difference is that the decoder now has to use the \textit{excess} in the checksum of $Y$ and compare it to its weight. In summary, when the edit is either a single deletion or insertion, one can synchronize $Y$ to $X$ with a simple zero-error algorithm that requires the encoder to transmit $\log (n+1)$ bits. No interaction is needed.

\section{Synchronizing from Multiple Deletions and Insertions} \label{sec:insdel}
\subsection{Only Deletions}
To illustrate the key ideas, we begin with the special case  where the sequence $Y$ is obtained by deleting $d>1$ bits from
$X$, where $d$ is $o(\tfrac{n}{\log n})$.  If the number of deletions is one, we know from Section~\ref{sec:single_ins_del} that $Y$ can be synchronized using a VT syndrome. The idea for $d >1$ is  to break down the synchronization problem into sub-problems, each containing only a single deletion. This is efficiently achieved through a divide-and-conquer strategy which uses interactive communication.

Consider the following example:
\be
  \begin{split}
 X  & = 1~0~0~1~1~0~0~\bm{\mathit{0}}~1~0~0~\un{1~0~1~0}~1~1~\bm{\mathit{0}}~1~1~0~0~1~1~0~\bm{\mathit{1}} \\
 Y  & = 1~0~0 ~ 1~1~0~0~ \,1~0~0~\un{1~0~1~0}~1~1~\, 1~1~0~0~1~1~0
  \end{split}
  \label{eq:del_example0}
\ee
where the deleted bits in $X$ are indicated by bold italics. It is assumed that the number of deletions $d=3$ is known to both the encoder and the decoder at the outset.
\begin{itemize}
\item In the first step, the encoder sends a few `\emph{anchor}' bits around the center of $X$ (underlined bits in \eqref{eq:del_example0}). The decoder tries to find a match for these anchor bits as close to the center of $Y$ as possible.  Here and in the remainder of the paper, finding a match for  a $k$-bit string $S$ around the center of an $l$-bit string  $Y$ ($l > k$) refers to the following: first check if $S$ matches the central substring of $Y$, defined as  the bits in locations $\{ \lfloor \frac{l}{2} - \frac{k}{2} +1\rfloor, \ldots ,  \lfloor \frac{l}{2} + \frac{k}{2}\rfloor \}$. If not, check if $S$ matches a substring of $Y$ located one position to the left/right of the center, and so on.

\item The decoder knows that the anchor bits correspond to positions $12$ to $15$ in $X$, but they align at positions $11$ to $14$ in $Y$.  Since the alignment  position is shifted to the left by one, the decoder infers that there is one deletion to the left of the anchor bits and two to the right, and conveys this information back to the encoder. (Recall that the lengths of $X$ and $Y$ are known. So the decoder knows that there are a total of three deletions as we have assumed  that all the edits are deletions.)

\item  The encoder sends the VT syndrome of the left half of $X$, using which the decoder corrects the single deletion in the left half of $Y$.
The encoder also sends a second set of anchor bits around the \emph{center of the right half} of $X$, as shown  below.
\end{itemize}
\ben
  \begin{split}
 X  & ={\color{gray}  1~0~0~1~1~0~0~\bm{\mathit{0}}~1~0~0~\un{1~0~1~0}}~1~1~\bm{\mathit{0}}~1~\un{1~0~0}~1~1~0~\bm{\mathit{1}} \\
 Y  & = {\color{gray} 1~0~0~1~1~0~0~ \,1~0~0~\un{1~0~1~0}}~1~1~\, 1~\un{1~0~0}~1~1~0
  \end{split}
  \label{eq:del_example1}
\een
\begin{itemize}
\item The decoder tries to find a match for these anchor bits as close to the center of  the right half of $Y$ as possible. The alignment position will indicate that there is one remaining deletion to the left of the anchor bits, and one to the right.

\item The encoder sends VT syndromes for the left and right halves of $X_r$, where $X_r$ is the substring consisting of bits in the right half of $X$.  Using the two sets of VT syndromes, the decoder corrects the remaining deletions.
\end{itemize}

The  example above can be generalized to a synchronization algorithm for the case where $Y$ is obtained from $X$ via $d$ deletions:
\begin{itemize}
  \item The encoder maintains an unresolved list $\mc{L}_X$, whose entries are  the yet-to-be-synchronized substrings of $X$. The list is initialized to be $\mc{L}_X = \{X\}$. The decoder maintains a corresponding list $\mc{L}_Y$, initialized to $\{ Y \}$.
  \item In each round, the encoder sends $m_a$ anchor bits around the center of each substring in $\mc{L}_X$ to the decoder, which tries to align these bits as close as possible to the center of the corresponding substring in $\mc{L}_Y$.   If a match is found, the aligned anchor bits split the substring into two pieces. For each of these pieces:
        \begin{itemize}
        \item[--]If the number of deletions is \textit{zero}, the piece has been synchronized.
        \item[--]If the number of deletions is \textit{one}, the decoder requests the VT syndrome of this piece for synchronization.
        \item[--]If the number of deletions is \textit{greater than one}, the decoder puts this piece in $\mc{L}_Y$. The encoder puts its corresponding piece in  $\mc{L}_X$.
        \end{itemize}
        
       If one or more of the anchor bits is among the deletions, the decoder may not be able to align the anchor bits. In this case, in the next round the decoder requests another set of $m_a$ anchor bits for the substring; this set is chosen adjacent to a previously sent set of anchor bits, as close to the center of the substring as possible. This process continues until the decoder is able to align a set of anchor bits for that substring.

 \item The process continues until $\mc{L}_Y$ (or $\mc{L}_X$) is empty.
 \end{itemize}

We now generalize the algorithm to handle a combination of insertions and deletions.

\subsection{Combination of Insertions and Deletions (Indels) } \label{subsec:indel_algorithm}

At the outset, both parties know only the lengths of $X$ and $Y$. Note that with indels, this  information does not reveal the total number of edits. For example, if the length of $Y$ is $n-1$, we can only infer that the number of deletions exceeds the number of insertions by one, but not exactly how many edits occured.

Consider the following example where the transformation from $X$ to $Y$ is via one deletion and one insertion. The deleted and inserted bits in $X$ and $Y$, respectively, are shown in bold italics.
\be
\begin{split}
{X} = & 1~1~0~1~1~0~0~0~\un{1~0~0}~1~0~1~\bm{\mathit{0}}~0~1~1~0 \\
{Y} =  & 1~1~0~1~1~0~0~0~\un{1~0~0}~1~0~1~\,0~1~1~0~\bm{\mathit{1}}
\end{split}
\label{eq:indel_example}
\ee
Since both the deletion and the insertion occur in the right half of $X$, the anchor bits around the center of $X$ will match exactly at the center of $Y$, as shown in \eqref{eq:indel_example}. When there are both insertions and deletions, the alignment position of the anchor bits only indicates the number of $\emph{net}$ deletions in the substrings to the left and right of the anchor bits. (The number of net deletions is the number of deletions minus the number of insertions.)
Thus, if the anchor bits indicate that a substring of $X$ has undergone zero net deletions, we need to check whether: a) the substring is perfectly synchronized, or  b) the alignment is due to an equal number of deletions and insertions $\ell$, for some $\ell \geq 1$.
To distinguish between these two alternatives, a hash comparison is used. 

Recall that a $k$-bit hash function applied to an $n$-bit binary string yields  a `sketch' or a compressed representation  of the string when $k<n$. For example, a simple $k$-bit hash function  is one that selects bits in randomly chosen positions $i_1, \ldots, i_k$.  Using such a hash, one could declare equal-length strings $A$ and $B$  identical if all the $k$ hash bits match. Note that every $k$-bit hash function  with $k < n$ has a non-zero probability of a hash collision, i.e., the event where two non-identical length-$n$ strings $A$ and $B$  hash to the same $k$ bits. 

In our synchronization algorithm,  whenever the anchor bits indicate that a substring has undergone zero net deletions, a hash comparison is performed to check whether the substring is synchronized. Similarly, if the anchor bits indicate that a substring has undergone one net deletion (or insertion), we hypothesize that it is due to a single bit edit and attempt to synchronize using a VT syndrome. A hash comparison is then used to then check if the substring is synchronized. If not, we infer that the one net deletion is due to $\ell$ deletions and $\ell-1$ insertions for some $\ell \geq 2$; hence further splitting is needed. 
The overall algorithm  works in a divide-and-conquer fashion, as described below.
\begin{itemize}
  \item The encoder maintains an unresolved list $\mc{L}_X$, whose entries  are the yet-to-be-synchronized substrings of $X$. This list is initialized to $\mc{L}_X = \{X\}$. The decoder maintains a corresponding list $\mc{L}_Y$, initialized to $\{ Y \}$.
  \item In each round, the encoder sends $m_a$ anchor bits around the center of each substring in $\mc{L}_X$. The decoder tries to align these bits  as close to the center of the corresponding substring in $\mc{L}_Y$ as possible. If a match is found, the aligned anchor bits split the substring into two pieces. For each of these pieces:
        \begin{itemize}
        \item[--]If the number of net deletions is \textit{zero}, the decoder requests  $m_h$ hash bits from the encoder to  check if the substring has been synchronized. If the hash bits all agree, it declares the piece  synchronized; otherwise the decoder adds the piece to $\mc{L}_Y$ (and instructs the decoder to add the corresponding piece to $\mc{L}_X$).
        \item[--] If the number of net deletions or insertions is \textit{one}, the decoder requests the VT syndrome of this piece as well as $m_h$ hash bits to verify synchronization. The decoder performs VT decoding followed by a hash comparison. If the hash bits all agree, it declares the piece synchronized; otherwise the decoder adds the piece to $\mc{L}_Y$ (and instructs the decoder to add the corresponding piece to $\mc{L}_X$).
        \item[--]If the number of net deletions or insertions is \textit{greater than one}, the decoder adds the piece to $\mc{L}_Y$ (and instructs the decoder to add the corresponding piece to $\mc{L}_X$). \end{itemize}
           If one or more of the anchor bits is among the edits, the decoder may not be able to align the anchor bits. In this case, in the next round the decoder requests another set of $m_a$ anchor bits for the substring;  this set is chosen adjacent to a previously sent set of anchor bits, as close to the center of the substring as possible. (This process continues until the decoder is able to align a set of anchor bits for that substring.)

  \item The process continues until $\mc{L}_Y$ (or $\mc{L}_X$) is empty.
\end{itemize}

The pseudocode for the algorithms at the encoder and decoder is given in the next page. For completeness and ease of analysis, we add the following rules to the synchronization procedure.
\begin{enumerate}

\item When the decoder receives $m_a$ anchor bits to be aligned within a substring of length $l$, it searches for a match within a window of length  $\kappa \sqrt{l}$ around the middle of its  substring, where  $\kappa \geq 1$ is a constant.

\item  If no matches for the anchor bits are found within this window, the decoder requests an additional set of anchor bits from a pre-arranged location, chosen as described above.

\item If multiple matches for the anchor are found within the window, the decoder chooses the match closest to the center of the substring. 

\item  Whenever an anchor needs to be sent for a piece whose length is less than $L (m_a + m_h)$,  the encoder just sends the piece in full. Here $L >1$ is a pre-specified constant.

\item Whenever the total number of bits transmitted in the course of the algorithm exceeds $\alpha n$ (for some pre-specified $\alpha \in (0,1)$), we terminate the algorithm and send the entire $X$ sequence.
\end{enumerate}

\begin{algorithm}
\caption{Synchronization Algorithm at the Encoder}
\label{alg:alg1}
\begin{algorithmic}[1]
 \STATE The encoder keeps a list $\mc{L}_X$ of unresolved substrings, which it initializes to $\mc{L}_X = \{ X \}$.
 \STATE In Round $1$:
          \IF{ length$(Y) = n$}
              	\STATE Send the hash of $X$.
 	\ELSIF { length$(Y) = n \pm 1$}
		\STATE Send both the VT syndrome and the  hash  of $X$
	\ELSE
		\STATE Send a set of $m_a$ anchor bits around the center of $X$
	\ENDIF
    \WHILE {$\mc{L}_X$ is non-empty}
      \STATE Receive from the decoder the instructions $I_s$ for all substrings $s \in \mc{L}_X$, and do the following for all $s\in \mc{L}_X$ in a single transmission:
      \FORALL{substrings $s\in \mc{L}_X$}
            \IF{$I_s=$ ``Matched''}
               \STATE Remove $s$ from $\mc{L}_X$.
      	   \ELSIF{$I_s=$ ``Anchor''}
               \STATE Send $m_a$ anchor bits around the center of $s$; if a set of anchor bits had already been sent for $s$ in the previous round, send a new set of $m_a$ anchor bits adjacent to a previously sent set, as close to the center of $s$ as possible.
	\ELSIF{$I_s=$ ``Split,x,y''}
               \STATE Split $s$ into two pieces $s_1,s_2$, using the previously sent anchor. Put $s_1,s_2$ into $\mc{L}_X$, and remove $s$ from  $\mc{L}_X$.
               	 \IF{x/y= ``Verify''}
              		\STATE Apply and send the hash of  $s_1/s_2$.
        		 \ELSIF{x/y= ``VT mode''}
             		  \STATE Send the VT syndrome and hash  for $s_1/s_2$.
		       	   \ELSIF{x/y= ``Anchor''}
               \STATE Send anchor bits around the center of $s_1/s_2$.
        		 \ENDIF
	 \ENDIF
    \ENDFOR
    \ENDWHILE
  \end{algorithmic}
\end{algorithm}
\begin{algorithm}
  \caption{Synchronization Algorithm at the Decoder}
  \label{alg:alg2}
  \begin{algorithmic}[1]
   \STATE The decoder keeps an  list $\mc{L}_Y$ of unresolved substrings, which is initialized to  $\mc{L}_Y = \{Y\}$. Define
   $I_Y$ to be ``Verify'' if length$(Y) =n$, ``VT Mode'' if  length$(Y) = n \pm 1$, and  ``Anchor'' otherwise.
   \WHILE{$\mc{L}_Y$ is non-empty}
       \STATE Read the instructions $I_s,s \in \mc{L}_Y$, and use them with the responses from the encoder  to decide the new set of instructions for each substring $s\in \mc{L}_Y$ as follows.
       \FORALL{substrings in $s \in \mc{L}_Y$}
           \IF{$I_s = $`Verify"}
               \STATE Compare the hash of $s$ with that sent by $X$. If the hashes match, add instruction ``Matched" for $s$, and remove $s$ from $\mc{L}_Y$; else add instruction ``Anchor" for $s$ and keep in $\mc{L}_Y$.
           \ELSIF{$I_s = $``VT mode"}
               \STATE Use the VT syndrome sent by $X$ to update the substring by deleting/inserting a single bit from $s$. Compare the hashes. If the hashes match, add instruction ``Matched" for $s$, and remove $s$ from $\mc{L}_Y$; else add instruction ``Anchor" for $s$ and keep in $\mc{L}_Y$.
           \ELSIF{$I_s = $  ``Anchor"}
               \STATE Try to find a substring near the center of  $s$ that matches with that sent by $X$. 
               
               If successful, split $s$ into two pieces, add each piece to   $\mc{L}_Y$, and remove $s$ from $\mc{L}_Y$.  Add the combined instruction ``Split,x,y", where ``Split'' is the instruction for $s$ and  x,y are the instructions for each of the two pieces of $s$. Each of x and y is one of  \{Verify, VT mode, Anchor\}, depending on whether the number of net deletions/insertions in the piece is 0,1, or a larger number. 
               
                If the anchor bits cannot be aligned,  request an adjacent set of anchor bits for $s$ by adding the instruction ``Anchor".
           \ENDIF
       \ENDFOR
       \STATE Send the new set of instructions to the encoder.
    \ENDWHILE
 \end{algorithmic}
\end{algorithm}

\emph{Choice of hash function}:   For our experiments in Section \ref{subsec:indel_exp}, we use the  $H_3$ universal class of hash functions, where the hash function $f: \{ 0,1 \}^n \to \{ 0,1\}^{m_h}$ of a $1 \times n$ binary string  $x$ is defined as
\be
f(x) = x \, \mbf{Q}
\label{eq:H3_hash}
\ee
where $\mbf{Q}$  is a binary $n \times m_h$ matrix with entries chosen i.i.d. Bernoulli$(\tfrac{1}{2})$, and the matrix multiplication is over GF($2$).  Such a hash function has a hash collision  probability of $2^{-m_h}$  whenever the compared strings are not identical \cite{CarterWeg77}.  We will choose the number of hash bits $m_h$ to be $c \log n$, where the constant $c$ determines the collision probability $n^{-c}$. Computing the $m_h$-bit hash in \eqref{eq:H3_hash} involves adding the rows of $\mbf{Q}$ that correspond to ones in $x$.  This computation requires $O(n)$ additions, each taking $O(\log n)$ bits.

In Section \ref{sec:sub_error}, we use a different hash function which serves as a Hamming distance estimator. Such a hash is useful when we are only interested in detecting whether the Hamming distance between the compared strings is greater than a specified threshold or not.

\emph{Computational Complexity}:  We can estimate the average-case complexity of the interactive algorithm, assuming a uniform distribution over the inputs and edit locations. When the number of edits $t=o(\tfrac{n}{\log n})$, the number of times anchor bits  are requested is also $o(\tfrac{n}{\log n})$.   The number of anchor bits $m_a$ sent each time  is  $c \log n$.   As discussed above, the hash computation for a length $n$ string requires $O(n)$ additions, each involving $O(\log n)$ bits. Computing the VT syndrome for a length $n$ string also requires $O(n)$ arithmetic operations, each involving  $O(\log n)$ bits.   Each bit of $X$/$Y$ is involved in a VT computation and a hash comparison only $O(1)$ times with high probability. The average computational complexity of the synchronization algorithm is therefore $O(n)$ arithmetic operations, with 
$O(\log n)$ bits of memory.

The following theorem characterizes the performance of the proposed synchronization algorithm when both the original string $X$ and the positions of the insertions and deletions are drawn uniformly at random.  For clarity, we set the number of anchor bits $m_a$ and the number of hash bits $m_h$  both equal to $c \log n$. We assume that the $c \log n$-bit hash  is generated from a universal class of hash functions \cite{CarterWeg77}, and thus has collision probability $\frac{1}{n^c}$. 

\begin{thm}
Let $X$ be a uniformly random length-$n$ binary sequence. Suppose that $Y$ is obtained from $X$ via $d$ deletions and $i$ insertions such that the total number of edits $t = (d+i) \sim o(\frac{n}{\log n})$, and the positions of the edits are uniformly random. Let the number of anchor bits  and hash bits (sent each time  they are requested) be $m_a = m_h =c \log n$, where $c  > 1.5$ is a constant.

{(a)} The probability that the algorithm fails to synchronize correctly, denoted by $P_e$, satisfies 
\[ P_e < \frac{t \log n}{n^c} + \frac{1}{n^{2(c-1)}}. \]

{(b)} Let $N_{1 \to 2} (t)$  and $N_{2 \to 1}(t)$ denote the number of bits transmitted by the encoder  and the decoder, respectively. Then for sufficiently large $n$:
\begin{align*}
 \expec N_{1 \to 2} (t) & < \left[(4c+2) t - (3c+1)\right] \log n,  \\
 \expec N_{2 \to 1} (t) & < 10(t-1) +1.
\end{align*}
\label{thm:indel}
\end{thm}

\emph{Remarks}:
\begin{enumerate}
\item The total communication required for synchronization is within a constant factor $(\approx 4c+2)$ of the fundamental  limit $t \log n$, despite the total number of edits being unknown to either party in the beginning.

\item The constant $c$ can be adjusted to trade-off between the communication error and the probability of error. In the proof of the theorem, the condition that $c > 1.5$ lets us analyze the effect of `bad' events such as anchor mismatch in a clean way. We expect that  the condition can be relaxed to $c > 1$ with a sharper analysis.  
\end{enumerate}
The proof of the theorem is given in Section \ref{subsec:indel_proof}.

\begin{table*}[t]
\caption{\small{ Average performance of the synchronization algorithm over $1000$ random binary $X$ sequences of length $n=10^6$. The edits consist of an equal number of deletions and insertions in random positions. }} \label{table:random_edits}
\vspace{-5pt}
\centering
\begin{tabular}{|c|c|c|c||c|c|c|c|}
\hline
No. of edits  & $m_a$ & \multicolumn{2}{|c||}{Bounds of Thm.\ref{thm:indel} ($\%$ of $n$)} & \multicolumn{3}{|c|}{Observed Values (Avg.) ($\%$ of $n$)} & \%  failed  \\
\cline{3-7}
 &  $=m_h$ & $\expec[N_{1 \to 2}] $ & $\expec[N_{2 \to 1}]$ & $N_{1 \to 2}$ & $N_{2 \to 1}$ & $N_{1 \to 2} + N_{2 \to 1}$ & trials  \\
\hline
100 &            &  0.793     &   0.0991      &   0.545     &   0.085       &  0.630     &   4.7 \\
500 & 10         &  3.981     &  0.4991      &   2.565     &    0.427      &  2.992     & 19   \\
1000 &            &   7.968    &    0.9991       &     4.989    &   0.853      &  5.842     & 34.4   \\
\hline
100 &    &   1.188       &    0.0991   &  0.905    &    0.082      &  0.987     &  0   \\
500 & 20   & 5.971       &    0.4991    &  4.338     &  0.410        & 4.748      &  0  \\
1000 &   &    11.1951    &      0.9991   &  8.481    &  0.817        &  9.298     &   0  \\
\hline
\end{tabular}
\vspace{-10pt}
\end{table*}

\subsection{Experimental Results} \label{subsec:indel_exp}
Table \ref{table:random_edits} compares the performance of the algorithm on uniformly random $X$ sequences with the bounds of Theorem \ref{thm:indel} as the number of edits $t$ is varied. The length of $X$ is fixed at $n=10^6$, and the edits consist of an equal number of deletions and insertions in random positions. Therefore the length of $Y$ is also $n=10^6$.  The $m_h$-bit hash is generated from the $H_3$ universal class, described in \eqref{eq:H3_hash}.

From Table \ref{table:random_edits}, we observe that the algorithm fails to synchronize reliably when the hash is only $10$ bits long. This  is consistent with the fact that the upper bound of Theorem \ref{thm:indel}(a) on the error probability exceeds $1$ for  $m_h=10$, even for $t=100$ edits. When $m_a=m_h=20$, there were no synchronization failures in any of the $1000$ trials.

The average number of bits sent in each direction is observed to be slightly less than the bound of Theorem \ref{thm:indel}(b). For example, when $Y$ differs from $X$ by $1000$ random edits, the algorithm synchronizes with overall communication that is less than  $10\%$ of the string length $n$.

\section{Synchronizing with a Limited Number of Rounds} \label{sec:limited_rounds}
Though the synchronization algorithm described in Section \ref{sec:insdel} has near-optimal rate and low computational complexity,   the number of rounds  of interaction grows as the logarithm of the number of edits $t$. To see this, recall that the algorithm  uses interaction to isolate $t$ substrings with exactly one insertion/deletion each. In each round, the number of substrings that $X$ has been divided into can at  most double, so at least $\log t$ rounds of interaction are required to isolate $t$ substrings with one edit each.

 In this section,  we show how  the  synchronization algorithm  can be modified to work with only one round of interaction. The reduction in the number of rounds comes at the expense of increased communication, which is characterized in Theorem \ref{thm:sing_round}.  Recall that the main purpose of interaction is to divide the sequence into substrings with only one deletion/insertion. These substrings are then synchronized using VT syndromes. To reduce the number of rounds, the idea is to divide $X$  into a number of equal-sized pieces such that most of the pieces are likely to contain either $0$ or $1$ edit. The encoder then sends anchor bits, hashes,  and VT syndromes for each of these pieces.   

 \begin{figure}[t]
\centering
\includegraphics[width=3.45in]{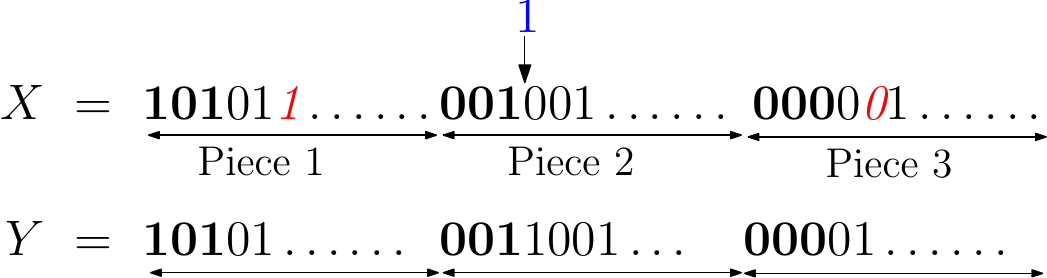}
\caption{\small{ $X$ is divided into equal-sized pieces. There is one deletion in the first piece of $X$, one insertion in the second piece, one deletion in the third piece etc.  Here the first three bits of each piece serve as anchor bits. The anchors allow the decoder to split $Y$ into pieces corresponding to those of $X$.}}
\vspace{-12pt}
 \label{fig:split_pieces}
\end{figure}

Let the length-$n$ string $X$ be divided into pieces of length $n^a$  bits for $a \in (0,1)$. There are $n^{\bar{a}}$ pieces, where $\bar{a}=1-a$.  The pieces are denoted by
$X_1,X_2,\ldots, X_{n^{\bar{a}}}$. The algorithm works as follows.

1)  For each piece $X_k$, $k =1 , \ldots, n^{\bar{a}}$, the encoder sends anchor bits, a hash, and the VT syndrome of $X_k$. The anchor of a  piece $X_k$ is a small number of bits indicating the beginning of $X_k$. The length $m_a$ of the anchor is $O(\log n)$. As illustrated in Fig. \ref{fig:split_pieces}, the anchors are used by the decoder to  split $Y$ into pieces corresponding to those of $X$.

2) The decoder sequentially attempts to synchronize the pieces.  For $k=1 , \ldots, n^{\bar{a}}$, it attempts to align the anchors for pieces $X_k$ and $X_{k+1}$ in order to determine the corresponding piece in $Y$, denoted $Y_k$. As in the previous algorithm, we try to align the anchor within a window of size approximately $\sqrt{n^a}$ around the center, since the length of each piece is $n^a$ bits. 
\begin{itemize}
\item[--] If the anchor  for either $X_k$  or $X_{k+1}$ cannot be aligned in $Y$, declare the $k$th piece to be unsynchronized.

\item[--] If $Y_k$ has length $n^a$, the piece  has undergone zero net edits. The decoder compares the hashes to check if the piece is synchronized. If the hashes disagree, it is declared unsynchronized.

\item[--] If $Y_k$ has length $n^a-1$ or $n^a +1$, the piece  has undergone one net edit. The decoder uses the VT syndrome to  perform VT decoding, and then uses the hashes to check if the piece is synchronized. If the hashes disagree, it is declared unsynchronized.

\item[--] If the lengths of $Y_k$ and $X_k$ differ by more than $1$, the number of edits is at least two. Declare the piece to be unsynchronized.
\end{itemize}

3) The decoder sends the status of each piece (synchronized/unsynchronized) back to the encoder.

4) The encoder sends the unsynchronized pieces to the decoder in full.

The algorithm thus consists of one complete round of interaction, followed by one transmission from the encoder to the decoder.
\begin{table*}[t]
\centering
\caption{\small{Average performance of the single-round algorithm  over $1000$ sequences for different values of $m_a=m_h$. Number of edits $=500$ ($d=i=250$).}}\label{table:single1}
\vspace{-6pt}
\begin{tabular}{|c|c|c|c|c|c|c|c|c|}
\hline
$m_a$  & \multicolumn{2}{|c|}{ $N_{1 \to 2} \, (\% \text{ of } n)$} & \multicolumn{2}{|c|}{ $N_{1 \to 2} + N_{2 \to 1} \, (\% \text{ of } n)$} &  \multicolumn{2}{|c|}{No. of pieces sent in full} & \multicolumn{2}{|c|}{\% failed trials}\\
\cline{2-9}
$= m_h$ &  $n = 10^6$ & $n = 10^7$ &   $n = 10^6$ & $n = 10^7$ & $n = 10^6$ & $n=10^7$ & $n =10^6$ & $n = 10^7$ \\
\hline
10 & 12.099 & 3.1279 &  12.199 & 3.2279 & 91.022 & 13.120 & 2.5 & 0.3 \\
15& 13.124 & 4.1189 & 13.224 & 4.2189 & 91.276 & 12.215 & 0.2 & 0 \\
20   & 14.147 & 5.1172 & 14.247 & 5.2172 & 91.499 & 12.050 & 0 & 0\\
25  & 15.192 & 6.1177 & 15.292 & 6.2177  & 91.957 & 12.096 & 0 & 0\\
30  & 16.237 & 7.1178 & 16.337 & 7.2178  & 92.404 & 12.104 & 0 & 0\\
\hline
\end{tabular}
\vspace{-5pt}
\end{table*}
\begin{table*}[t]
\centering
\caption{\small{Average performance of the single-round algorithm  over $1000$ sequences as the number of edits is varied. The number of anchor and hash bits is fixed at
$m_a=m_h=20$.}}\label{table:single2}
\vspace{-5pt}
\begin{tabular}{|c|c|c|c|c|c|c|}
\hline
Number of    & \multicolumn{2}{|c|}{ Bound of Thm.\ref{thm:sing_round} for} & \multicolumn{2}{|c|}{Average observed} & \multicolumn{2}{|c|}{Average no. of pieces} \\
 edits  & \multicolumn{2}{|c|}{$\expec N_{1 \to 2} + N_{2 \to 1} (\% \text{ of } n)$} & \multicolumn{2}{|c|}{$N_{1 \to 2} + N_{2 \to 1}\, (\% \text{ of } n)$} & \multicolumn{2}{|c|}{sent in full}\\
\cline{2-7}
& $n = 10^6$ & $n = 10^7$ &  $n = 10^6$ &  $n = 10^7$ &  $n = 10^6$ &  $n = 10^7$\\
\hline
20  & 5.197  &  5.1049 & 5.116  & 5.0969 & 0.192 & 0.02 \\
50 & 5.422  &  5.1180  & 5.222  & 5.0980 & 1.2520 & 0.125  \\
100  & 5.997 & 5.1417 & 5.559  & 5.1012 & 4.6270 & 0.445 \\
300  & 10.797 & 5.2617 & 8.853 & 5.1409 & 37.563 & 4.414 \\
500  & 19.597 & 5.4217 & 14.247 & 5.2172 & 91.4990 & 12.05 \\
\hline
\end{tabular}
\vspace{-5pt}
\end{table*}
The following theorem characterizes its performance.
\begin{thm}
Suppose that $Y$ is obtained from a uniformly random length-$n$ binary sequence $X$ via $t = n^b$ indel edits, where $b \in (0,1)$ and the locations of the edits are uniformly random.   Let $n^a$ be  the size of each piece in the single-round algorithm above, with $a \in (0,1)$. Let the number of anchor bits  and hash bits per  piece be equal to $m_a = c_a \log n$ and  $m_h = c_h \log n$, respectively.  Then for  $b < 1-a$, the  one-round algorithm has the following properties.

{(a)} The probability of error, i.e.,  the probability that the algorithm fails to synchronize correctly is less than $\frac{1}{n^{c_h+a -1 }}$.

{(b)}  For $c_a > (1 +\frac{a}{2})$, the total number of bits transmitted by the encoder, denoted $N_{1 \to 2}$ satisfies 
\be \begin{split} 
\expec N_{1 \to 2}  <  &   \Big( ( c_a + c_h + a) n^{1-a} \log n  \  +  \ \tfrac{1}{2} n^{2a+2b-1} \\ 
& \  \  +   2c_a n^b \log n\Big)(1 + o(1)).  \end{split} \label{eq:eN21}  \ee
The number of bits transmitted from the decoder to the encoder is deterministic and equals $n^{1-a}$.
\label{thm:sing_round}
\end{thm}

The proof of the theorem is given in Section \ref{subsec:sing_round_proof}.

\emph{Remarks}:

1)  As $n \to \infty$, the expected communication  is minimized when the exponents of the first two terms in  \eqref{eq:eN21} are balanced.  This happens when
\[ 1-a = 2a +2b -1.  \]
Therefore the optimal segment parameter $a$ for a given number of edits $n^b$ is $\tfrac{2}{3}\bar{b}$.  With this value, the total number of bits transmitted is
$\Theta (n^{(1+2b)/3} \log n )$.

As an  example, suppose that the number of edits $t=\sqrt{n}$. Then $b=0.5$, and the optimal value of $a=\tfrac{1}{3}$. With this choice of $a$ and  
$m_a = m_h =  2 \log n$,   the bound of Theorem \ref{thm:sing_round} yields
\[ \expec N_{1 \to 2} <  \left( \frac{13}{3} n^{2/3} \log n + \frac{1}{2} n^{2/3} + 4 n^{1/2} \log n\right)(1 + o(1)). \]
Further, $N_{2 \to 1} = n^{2/3}$,  and the probability of synchronization error is bounded by $n^{-4/3}$.

2) In general, we may not know the number of edits beforehand. For a given segment size $n^a$, the algorithm can handle up to $n^{\bar{a}}$ random edits by communicating $o(n)$ bits. This is because the algorithm is effective when most blocks have zero or one edits, which is true when $b < \bar{a}$. If the number of edits is larger than $n^{\bar{a}}$, it is cheaper for the encoder to send the entire $X$ sequence.

3) The original interactive algorithm and the one-round algorithm represent two extreme points of the trade-off between the number of rounds and the total communication rate required for synchronization. It is possible to interpolate between the two and design an algorithm that uses at most $k$ rounds of interaction for any constant $k$.

\subsection{Experimental Results}

The single-round algorithm was tested on uniformly random binary $X$ sequences of  length $n = 10^6$ and $n = 10^7$.  Each piece of $X$ was chosen to be $1000$ bits long. Therefore  $X$ was divided into 1000 pieces for $n=10^6$, and 10000 pieces for   $n=10^7$, corresponding to section parameter values $a=0.5$ and $a=0.429$, respectively.

$Y$ was obtained from $X$ via $t=500$ edits, with an equal number of deletions and insertions. Table \ref{table:single1} shows the average performance over $1000$ trials as $m_a$ and $m_h$ are varied, with $m_a=m_h$. We observe that (with $m_a=m_h=20$) we have reliable synchronization from $t=500$ edits with a communication  rate of  $14.2 \%$ and $5.2 \%$  for $n=10^6$, and $n=10^7$, respectively. In comparison, Table \ref{table:random_edits} shows that the multi-round algorithm needs a rate of only $4.75 \%$ for  $n=10^6$ for synchronizing from the the same number of edits. This difference in the communication required for synchronization reflects the cost of allowing only one round of interaction.

Table \ref{table:single1} also lists the number of pieces that need to be sent in full by the encoder in the second step. These are the pieces that either contain more than one edit, or contain an edit in  one of the anchor bits. Observe that the fraction of pieces that remain unsynchronized at the end of the first step is around $9.1 \%$ for  $n=10^6$, and only $0.13 \%$ for $n=10^7$. This is because the $t=500$ edits are uniformly distributed across $1000$ pieces in the first case, while the edits are distributed across $10,000$ pieces in the second case.  Therefore, the bits sent in the second step of the algorithm form the dominant portion of $N_{1 \to 2}$ for $n=10^6$, while the bits sent in the first step dominate $N_{1 \to 2}$ for $n=10^7$.

 Table \ref{table:single2}  compares the observed performance of the single-round algorithm with the upper bound of Theorem \ref{thm:sing_round} as the number of edits is varied, with the number of hash and anchor bits per piece fixed to be $20$.  The edits consist of an equal number of deletions and insertions. As in the previous experiment, $X$ is divided into pieces of $1000$ bits each. Observe that the total number of bits sent begins to grow with the number of edits only when the number of edits is large enough for $N_{1 \to 2}$ to have a significant contribution from the pieces sent in the second step.

We also compared the performance of the single-round algorithm to rsync, which also uses only one round of interaction. For $X$ and $Y$ differing by $500$ random edits, the amount of data required required to be sent by rsync (on average)  was $133 \%$ of the file size for both $n=10^6$ and $n=10^7$. Since the communication required by rsync is greater than the file size in both cases, sending  $X$ in full is the better option, which is invoked by most implementations of rsync. 

 In rsync, $Y$ is split into pieces and hashes for each piece are sent to the encoder. Pieces for which a match is not found are then sent in full with assembly instructions. When the edits are uniformly spread across the file, only a few pieces of $Y$ will have a match in $X$, thereby causing rsync to send a large part of $X$ (along with assembly instructions) in the second step. Thus rsync saves bandwidth only when the edits  are restricted to a few small parts of a large file rather than being spread throughout the file.

\section{Synchronizing from Bursty Edits}\label{sec:bursts}
Burst deletions and insertions can be a major source of mis-synchronization in practical applications as editing often involves modifying chunks of a file rather than  isolated bits. Recall that the algorithm described in Section \ref{sec:insdel} seeks to divide the original string into pieces with one insertion/deletion each, and uses VT syndromes to synchronize each piece. It is shown in Section \ref{subsec:indel_proof} that the expected number of times that anchor bits are requested is approximately $2t$ when the locations of the $t$ edits are uniformly random. However, when there is a burst of deletions or insertions, attempting to isolate substring with exactly one edit is inefficient, and the number of bits sent by the algorithm in each direction grows by a factor of $\log n$.

In this section, we  first describe a method to efficiently synchronize from a \emph{single} burst (of either deletions or insertions) of known length, and then generalize the algorithm of Section \ref{sec:insdel} to efficiently handle multiple burst edits of varying lengths.

\subsection{Single Burst} \label{subsec:sing_burst}
Suppose that  $Y$ is obtained from the length-$n$ string $X$ by deleting or inserting a single burst of $B$ bits. We allow $B$ to be a function of $n$, e.g. $B=\sqrt{n}$, or even $B=\alpha n$ for some small $\alpha >0$.  A lower bound on the number of bits required for synchronization can be obtained by assuming the encoder knows the exact location of the burst deletion. Then it has to send  two pieces of information to  the decoder: a) the location of the starting position of the burst, and b) the actual bits that were deleted. Thus the number of bits required, denoted $N_{burst}(B)$, can be bounded from below as
\be
N_{burst}(B) > B+ \log n,
\label{eq:burst_lb}
\ee

The goal is to develop a synchronization algorithm whose performance is close to the lower bound of \eqref{eq:burst_lb}. Let us divide each of $X$ and $Y$ into $B$ substrings as follows. For $k=1,\ldots,B$, the substrings $X^k$ and $Y^k$ are defined as
\be
\begin{split}
X^k & = (x_k, x_{B+k},x_{2B+k},\ldots), \\
 Y^k & = (y_k, y_{B+k},y_{2B+k},\ldots).
\end{split}
\label{eq:bit_planes}
\ee
Consider the following example where $X$ undergoes a burst deletion of $B=3$ bits (shown in red italics): 
\be X = 10 {\color{red} \emph{011}}100100011,  \quad Y = 10100100011  \label{eq:single_burst}. \ee
The three substrings formed according to \eqref{eq:bit_planes} with $B=3$ are
\be
\begin{split}
X^1 = 1 {\color{red} \emph{1}}001, \  \ & X^2=0{\color{red} \emph{1}}001, \  \ X^3={\color{red} \emph{0}}110, \\
Y^1= 1001, \  \ & Y^2=0001, \  \ Y^3=110.
\end{split}
\label{eq:bp_example}
\ee

 Observe that each of the substrings $X^k$ undergoes exactly one deletion to yield $Y^k$.  Whenever we have a single burst deletion (insertion) of $B$ bits, and divide $X$ and $Y$ into $B$ substrings as in \eqref{eq:bit_planes},  $X^k$ and $Y^k$ differ by exactly one deletion (insertion)  for  $k=1,\ldots,B$. Moreover, the positions of the single bit deletions in the $B$ substrings $\{ X^k \}_{k=1}^B$ are \emph{highly correlated}. In particular, if the deletion in substring $X^1$ is at position $j$, then the deletion in the other substrings is either at position $j$ or $j-1$. In the example
\eqref{eq:single_burst}, the second bit of $X^1$ and $X^2$, and the first bit of $X^3$ are deleted. More generally, the following property  can be verified.

\textbf{ \emph{Burst-Edit Property}}: Let $Y$ be obtained from $X$ through a single burst deletion (insertion) of length $B$, and let substrings $X^k$ be defined as in \eqref{eq:bit_planes} for $k=1, \ldots,B$.  Then if $p_k$ denotes the position of the deletion (insertion) in substring $X^k$, we have:
\[ p_k \geq p_{k+1}, \text{ for } k=1, \ldots, (B-1), \quad \text{ and  } \quad  p_1 \leq \, p_B + 1. \]
In other words, the position of the edit is non-increasing and can decrease at most once as we enumerate the substrings $X^k$ from $k=1$ to $k=B$.

This property suggests a  synchronization algorithm of the following form:
\begin{enumerate}
\item The encoder sends the VT syndrome of substring $X^1$. (Requires $\log(1 +n/B)$ bits.)

\item The decoder synchronizes $Y^1$ to $X^1$, and sends the position $j$ of the edit back to the encoder.  (Requires $\log(n/B)$ bits.)

\item For  $k=2,\ldots,B$, the encoder sends the  bits in positions $(j-1)$ and $j$ of $X^k$. (Requires $2(B-1)$ bits.)

The decoder reconstructs each $X^k$ by inserting/deleting the received bits in positions $(j-1)$ and $j$ of $Y^k$. 
\end{enumerate}

In the second step above, we have implicitly assumed that by correcting the single deletion/insertion in $Y^1$, the decoder can determine the exact position of the deletion in $X^1$. However, this may not always be possible. This is because the VT code always inserts a deleted bit (or removes an inserted bit) either at the beginning or the end of the \emph{run} containing it.
In the example in \eqref{eq:bp_example}, after synchronizing $Y^1$ to $X^1$, the decoder can only conclude that a bit was deleted in $X^1$ in either the first or second position.

 To address this issue, we modify the first two steps as follows. In the first step, the encoder sends the VT syndromes of both the first and last substrings, i.e., of $X^1$ and $X^B$. Suppose that the single edit in $X^1$ occurred in the run spanning positions $j_1$ to $l_1$, and the edit in $X^B$ occurred in the run spanning positions $j_B$ to $l_B$. Then, the burst-edit property implies that in the final step, the encoder only needs to send the bits in positions $j^*$ to $l^*$ of each substring $X^k$, where
\be
j^*= \max \{j_1-1, j_B \},  \quad l^*=\min \{ l_1, l_B +1 \}.
\label{eq:jsls}
\ee
We note that for \emph{any} substring $X^k$, $j^*$ is the first possible location of the edit, and $l^*$ is the last possible location of the edit.
The final algorithm for  exact synchronization from a single burst deletion/insertion is summarized as follows.

\emph{Single Burst Algorithm}:
\begin{enumerate}
\item The encoder sends the VT syndrome of substrings $X^1$ and $X^B$. (Requires $2\log(1 +n/B)$ bits.)

\item The decoder synchronizes $Y^1$ to $X^1$, and $Y^B$ to $X^B$. For each of the two substrings, the decoder sends back the index of the run containing the edit.   (Requires $2\log(n/B)$ bits.)

\item For  $k=2,\ldots,B-2$, the encoder sends bits in positions $j^*$ through $l^*$ of $X^k$. (Requires $(l^* - j^* +1)(B-2)$ bits.)

The decoder reconstructs each $X^k$ by inserting/deleting the received bits in positions $j^*$ through $l^*$ of $Y^k$.
\end{enumerate}

 We note that the algorithm does not make any errors. The following theorem shows that  when $X$ is a uniformly random  sequence, the expected number of bits required to synchronize is  within a small factor of the lower bound in \eqref{eq:burst_lb}.

\begin{thm}
Let $X$ be a uniformly random binary sequence of  length $n$. Let $Y$ be obtained via a single  burst of deletions (or insertions) of length $B$, with the starting location of the burst being uniformly random. Then for sufficiently large $n$, the expected number of bits sent by the encoder in the single burst algorithm satisfies
\ben
\begin{split} \expec N_{1 \to 2}  >  \, &  2\log (1 +n/B)  + (2-\tfrac{1}{B}) (B-2)  \\ 
 \expec N_{1 \to 2} \leq & \, 2 \log (1 +n/B) +  3 (B-2). \end{split} \een
The expected number of bits sent by the decoder is $2 \log(n/(2B))$.
\label{thm:single_burst}
\end{thm}
The proof of the Theorem is given in Section \ref{subsec:sing_burst_proof}.

\subsection{Multiple Bursts} \label{subsec:mult_bursts}
We can now modify the original synchronization algorithm to handle multiple edits, some of which occur in isolation and others in bursts of varying lengths. The idea is to use the anchor bits together with interaction to identify pieces of the string with either one deletion/insertion or one burst deletion/insertion. Since a burst consists of a number of adjacent deletions/insertions, it can be detected by examining the offset indicated by the anchor bits. In particular, if the offset for a particular piece of the string is a large value $B$ that is unchanged after a few rounds, we hypothesize a burst edit of length $B$.  This is because the isolated edits are likely to be spread across $X$, causing the offset to change within a few rounds.

We include the following ``guess-and-check" mechanism in the original synchronization algorithm: When the number of net deletions (or insertions) in a substring is greater than a  specified threshold $B_0$, and does not change after a certain number of rounds (say $T_{burst}$), we hypothesize that a burst deletion (or insertion) has occurred, and invoke the single burst algorithm of Section \ref{subsec:sing_burst}. In other words, we  correct the substring assuming a burst occurred and then use  hashing to verify the results of the correction. If the hashes agree, we declare that the substring has been synchronized correctly, otherwise we infer that the deletions (or insertions) did not occur in a burst, and continue to split the substring. The value of $T_{burst}$ can be adjusted to trade-off between the number of rounds and  the amount of total communication.

\subsection{Experimental Results}
\emph{Case 1: Single Bursts}.
The single-burst algorithm was tested on uniformly random $X$ sequences of length $n = 10^6$ and $n = 10^7$ with a single burst of deletions introduced at a random position. Table \ref{table:single_bursts} shows the average number of bits (over 1000 trials) transmitted from the encoder to the decoder for various burst lengths.
\begin{table}[t]
\centering
\caption{\small{Performance of single-burst algorithm over $1000$ trials}}\label{table:single_bursts}
\vspace{-5pt}
\begin{tabular}{|c|c|c|c|}
\hline
Length   & Thm. \ref{thm:single_burst} upper &  Avg. $N_{1 \to 2}$  & Avg. $N_{1 \to 2}$  \\
of burst &   bound on $\expec N_{1 \to 2}$  &  for $n = 10^6$ & for $n = 10^7$ \\
\hline
$10^2$ & 294 & 290 & 264.4\\
$10^3$ & 2994 & 2680 & 2632\\
$10^4$ & 29994 & 26110 & 26270\\
$10^5$ & 299994 & 257000 & 260200\\
\hline
\end{tabular}
\vspace{-12pt}
\end{table}

\emph{Case 2: Multiple bursts and isolated edits}.
The  algorithm of Section \ref{subsec:mult_bursts} was  tested on  a combination of isolated edits and multiple bursts of varying length. Starting with uniformly  random binary $X$ sequences of length $n = 10^6$,  $Y$ was generated via a few burst edits followed by a few isolated edits. The length of each burst was a random integer chosen uniformly in the interval  $[80,200]$. Each isolated/burst edit was equally likely to be  deletion or an insertion, and the locations of the edits were randomly chosen. Table \ref{table:multi_bursts} shows the average performance over $1000$ trials with $m_a=m_h=20$ bits. We set $T_{burst} = 2$: whenever there the offset of a piece is unchanged and large ($>50$) for two consecutive rounds, the burst mode is invoked.
\begin{table}[t]
\centering
\caption{\small{Performance of the algorithm on  a combination of multiple bursts and isolated edits. The length of $X$ is
$n =10^6$.}}\label{table:multi_bursts}
\vspace{-5pt}
\begin{tabular}{|c|c|c|c|c|c|}
\hline
No. of & No. of & Avg. & Avg. & Avg.  $N_{1 \to 2}$\\
bursts &  isolated edits & $N_{1 \to 2}$ &  $N_{2 \to 1}$ &  + $N_{2 \to 1}$\\
\hline
3 & 10    & 2139.1 & 242.6 &  2381.7 \\
3 & 15    & 2488.6 & 290.8 &  2779.4 \\
4 & 10    & 2623.6 & 296.9 &  2920.5 \\
4 & 15    & 2956.2 & 346.8 &  3303.0\\
5 & 10    & 3100.8 & 347.2 &  3448.0\\
5 & 15    & 3436.3 & 400.6 &  3836.9 \\
5 & 50    & 5889.7 & 756.3 &  6646.0 \\
\hline
\end{tabular}
\end{table}
We observe that the algorithm synchronizes from a combination of $50$ isolated edits and $5$ burst edits with lengths uniformly distributed in $[80,200]$  with a communication rate smaller than $1 \%$. This indicates that having prior information about the nature of the edits---an upper bound on the size of the bursts, for example---can  lead to significant savings in the communication required for synchronization.

\section{Correcting substitution edits}  \label{sec:sub_error}

In many practical applications, the edits are a combination of substitutions, deletions, and insertions.
The goal in this section is to equip the synchronization algorithm of Section \ref{sec:insdel} to handle substitution errors in addition to deletions and insertions. The approach is to first correct a large fraction of the deletions and insertions so that the decoder has a length-$n$ sequence $\hat{X}$ that is within a target Hamming distance $d_0$ of $X$. 
Perfect synchronization can then be achieved  by sending the  syndrome of $X$ with respect to a linear error-correcting code (e.g. Reed-Solomon or LDPC code) that can correct $d_0$ substitution errors  \cite{Wyner74,Orlitsky93,PradhanR03}. 

Since synchronizing two equal-length sequences with Hamming distance bounded by $d_0$ is a well-understood problem,  we focus here on the first step, i.e., the task of synchronizing $Y$ to  within a target Hamming distance of $X$. For this, we use locality-sensitive hashing, where the probability of hash collision is related to the distance between the two strings being compared. We use the sketching technique of Kushilevitz et al. \cite{Kushil98} to obtain a Hamming distance estimator which will serve as a locality-sensitive hash. In Section \ref{subsec:dist_sync}, this hash is used in the interactive algorithm of Section \ref{sec:insdel} to synchronize $Y$ to within a target Hamming distance of $X$.

\subsection{Estimating the Hamming Distance}
Suppose Alice and Bob have length-$n$ binary sequences $x$ and $y$, respectively.  Alice  sends  $m_h < n$ bits in order for Bob to estimate the Hamming distance $d_H(x,y)$ between $x$ and $y$. Define the hash function  $g: \{ 0,1 \}^n \to \{ 0,1\}^{m_h} $  as
\be
g(x) = x \mbf{R}
\label{eq:bin_sketch}
\ee
where $\mbf{R}$  is a binary $n \times m_h$ matrix with entries chosen i.i.d Bernoulli$(\tfrac{\kappa}{2n})$, and the matrix multiplication is over GF(2). $\kappa$ is a constant that controls the accuracy of the distance estimate, and will be specified later.  Define the function ${Z}$ as
\be
{Z} (x,y) =  g(x) \oplus  g(y)
\ee
where $\oplus$ denotes modulo-two addition.  Let
\[ {Z} (x,y) = (Z_1(x,y), \, Z_2(x,y), \ldots, Z_{m_h}(x,y)). \]
$Z_i(x,y)$ is the indicator function $1_{\{h_i(x) \neq h_i(y)\}}$ for $i=1, \ldots, m_h$. We have
\be
\begin{split}
P(Z_i(x,y)=1) & = P \left(  \sum_{l=1}^n x_l R_{li} \oplus  \sum_{l=1}^n  y_l R_{li} =1 \right)  \\
& =  P \left(  \sum_{l: x_l \neq y_l} R_{li}   =1 \right)
\end{split}
\label{eq:hash1_prob}
\ee
where the summations denote modulo-two addition. Since the matrix entries $\{ R_{li} \}$ are i.i.d. Bernoulli$(\tfrac{\kappa}{2n})$, it is easily seen (e.g., via induction over the summands in the \eqref{eq:hash1_prob}) that
\be
P(Z_i(x,y)=1) =  p \triangleq  \frac{1}{2} \left(  1 -  \left(1 - \frac{\kappa}{n}\right)^{d_H(x,y)} \right), 
\label{eq:Zi_dist}
\ee
for $i=1, \ldots,m_h$. Further, for any pair $(x,y)$, the random variables $Z_i(x,y)$ are i.i.d. Bernoulli with the distribution  given in \eqref{eq:Zi_dist}. This because the random matrix entries
$\{ R_{li} \}$ are i.i.d. for $1 \leq i \leq m_h$ and $1 \leq l \leq n$. The empirical average  of the entries of $Z(x,y)$, given by
\be
\bar{Z} (x,y) = \frac{1}{m_h} \sum_{i=1}^{m_h}  Z_i(x,y)
\label{eq:barZ}
\ee
has expected value equal to the right side of \eqref{eq:Zi_dist}. For large $m_h$, $\bar{Z}$ will concentrate around its expected value, and can hence be used to estimate the Hamming distance. Inverting \eqref{eq:Zi_dist}, we obtain the Hamming distance estimator
\be
\hat{d}_H(x,y) = \left\{
\begin{array}{ll}
\frac{\ln(1 - 2 \bar{Z})}{\ln(1 - \kappa/{n})} &  \text { if }  \bar{Z} \leq  \frac{1}{2} \left(  1 -  \left(1 - \frac{\kappa}{n}\right)^n \right)  \\
n & \text { otherwise}
\end{array}
\right.
\label{eq:ham_estimator}
\ee
We note that a related but different sketching technique for estimating the Hamming distance was used in  \cite{CormodePSV00}.
\begin{prop}
Consider any pair of sequences $x,y \in \{0,1\}^n$ with Hamming distance $d_H(x,y)$. Let $p$ be as defined in \eqref{eq:Zi_dist}. For $\delta \in (0, \tfrac{1}{2} -p)$, the following bounds hold for the Hamming distance estimator  in \eqref{eq:ham_estimator}.
\begin{align}
& P\Big( \frac{\hat{d}_H(x,y)}{n}  > \frac{d_H(x,y)}{n} + \frac{2 \delta}{\kappa (1-2p)}  + \frac{2 \delta^2}{\kappa (1-2p)^2}  \Big) \nonumber \\ 
& < e^{-2m_h \delta^2}, \label{eq:sketch_ub}\\
& P\Big( \frac{\hat{d}_H(x,y)}{n} <  \frac{d_H(x,y)}{n} - \frac{2 \delta}{\kappa (1-2p)}  +  \frac{2 \delta^2}{\kappa (1-2p)^2}  \Big) \nonumber \\
& < e^{-2m_h \delta^2}. \label{eq:sketch_lb}
\end{align}
\label{prop:sketch_chernoff}
\end{prop}
\proof In Appendix \ref{app:sketch_chernoff_proof}.

Using the approximation
\be
(1-2p) =  \left(1 - \frac{\kappa}{n}\right)^{d_H(x,y)} \approx \exp(-\kappa \tfrac{d_H(x,y)}{n})
\ee
for large $n$ in \eqref{eq:sketch_ub} and \eqref{eq:sketch_lb}, Proposition \ref{prop:sketch_chernoff}  implies  that  for small values of $\delta$,  the (normalized) Hamming distance estimate $\tfrac{1}{n} \hat{d}_H(x,y) $ lies in the interval
\be
\frac{d_H(x,y)}{n} \  \pm  \ \frac{2.2  \exp\left(\kappa \tfrac{d_H(x,y)}{n}\right) \delta}{\kappa}
\label{eq:conf_int}
\ee
with probability at least $1 - 2e^{-2m_h\delta^2}$. (The constant $2.2$ in \eqref{eq:conf_int} can be replaced by any number greater than $2$.)

 In the synchronization algorithm, we will use the distance estimator to resolve questions of the form  ``is  the distance $\tfrac{1}{n} d_H(x,y)$ is less than $d_0$?".
 The parameter $\kappa$ used to define the hashing matrix in \eqref{eq:bin_sketch} can be fixed using \eqref{eq:conf_int} as a guide. Setting $\kappa = 1/d_0$ implies that that the estimated distance $\tfrac{1}{n} \hat{d}_H(x,y)$ lies in the interval
\be
 \frac{1}{n}  d_H(x,y)  \  \pm \  2.2  \exp\left(\tfrac{1}{d_0} \tfrac{d_H(x,y)}{n}\right) d_0 \delta
\label{eq:dhat_lb}
\ee
with probability at least $1 - 2e^{-2m_h\delta^2}$. For example, if the actual distance $\tfrac{1}{n} d_H(x,y) = d_0$, the bound in \eqref{eq:dhat_lb} becomes
\be
 d_0 (1 -   5  \delta)  \ < \ \frac{\hat{d}_H(x,y)}{n}     \  <  \  d_0 (1 +   5  \delta).
\ee

\begin{table*}[t]
\caption{\small{ Average performance of the synchronization algorithm with the distance estimator hash. Length of $X$ is $n=10^6$. $Y$ was generated via $10$ deletions, $10$ insertions, and $100$ substitutions. }} \label{table:sub_edits}
\vspace{-5pt}
\begin{center}
\begin{tabular}{|c|c|c|c|c|c|c|}
\hline
Hash & Hash  & Initial (norm.) & Final  (norm.) & Avg. $N_{1 \to 2}$  & Avg. $N_{2 \to 1}$ & Avg $N_{1 \to 2} + N_{2 \to 1}$ \\
 length & type & Hamm. Dist. & Hamm. Dist. & ($\%$ of $n$) & ($\%$ of $n$)  &  ($\%$ of $n$) \\
\hline
10 & $H_3$ & $0. 3667 $ & $6.17 \times 10^{-4}$  & 2.937  & 0.5710  & 3.508 \\
\hline
& $\hat{d}_H$ & $0.3667 $ & $2.35 \times 10^{-2}$ & 0.208  & 0.0291 & 0.237 \\
\hline
20 & $H_3$ & $0. 3622 $ & 0  & 4.436  & 0.5314  & 4.968 \\
\hline
& $\hat{d}_H$ & $0.3622 $ & $2.2 \times 10^{-3}$ & 0.446  &  0.0423 & 0.488  \\
\hline
40 & $H_3$ & $0. 3653 $ & 0  & 7.413   & 0.5302  & 7.943 \\
\hline
& $\hat{d}_H$ & $0.3653 $ & $3.47 \times 10^{-4}$  & 0.798 & 0.0466 &  0.845 \\
\hline
\end{tabular}
\end{center}
\vspace{-10pt}
\end{table*}

\subsection{Synchronizing $Y$ to within a target Hamming distance of $X$} \label{subsec:dist_sync}

We  use the Hamming distance estimator as a hash in the synchronization algorithm of Section \ref{subsec:indel_algorithm}. The idea is to fix a constant $d_0 \in (0,1)$,  and declare synchronization between two substrings whenever the normalized Hamming distance estimate between them is less than $d_0$. The parameter $\kappa$ used to define the hash function $h$ in \eqref{eq:bin_sketch} is set equal to $1/d_0$.

The synchronization algorithm of Section \ref{subsec:indel_algorithm} is modified as follows. Whenever a hash is requested by the decoder, the encoder sends $g(x)$. The decoder computes $Z=g(x) \oplus g(y)$ and
$\hat{d}_H(x,y)$ as in \eqref{eq:barZ}.   (Here, $x$ and $y$ denote the equal-length sequences at the encoder and decoder, which are to be compared.)  If the normalized $\hat{d}_H(x,y)$ is less than $d_0$, declare synchronization; else put this piece in $\mc{L}_Y$ (and correspondingly in $\mc{L}_X$). The rest of the synchronization algorithm remains the same.

After the final step, the encoder may estimate the Hamming distance between $X$ and the synchronized version of $Y$ using another hash $g(y)$. Perfect synchronization can then be achieved by using the syndromes of a linear code of appropriate rate. We note that the distance estimator can also be used in the  algorithms described in Sections \ref{sec:limited_rounds} (limited rounds) and \ref{subsec:mult_bursts}  (multiple bursty edits) to achieve synchronization within a target Hamming distance.

Besides isolating the substitution edits, we note that a distance-sensitive hash also saves communication whenever a deletion and insertion occur close to one another giving rise to equal-length substrings with small normalized Hamming distance between them.

\subsection{Experimental Results}
 Table \ref{table:sub_edits} compares the performance of the synchronization algorithm  with the Hamming distance estimator hash for uniformly randomly $X$ of length $n = 10^6$. To clearly understand the effect of substitution edits,  $Y$ was generated from $X$ via $10$ deletions, $10$ insertions, and $100$ substitutions at randomly chosen locations. The number of anchor bits was fixed to be $m_a=10$, while the number of bits used for the hash/distance estimator was varied as $m_h = 10, 20, 40$. The table shows the average performance over $1000$ trials.

 The  parameter of the distance estimator was set to be $\kappa=50$, and we declare synchronization between two substrings if the estimated (normalized) Hamming distance is less than $d_0=0.02$. Table \ref{table:sub_edits} also shows the performance using  a standard universal hash $H_3$, described in \eqref{eq:H3_hash}. In each case, if the length of the two strings being compared was less than $m_h$, the encoder sends its string in full to the decoder. This is the reason the $H_3$ hash
is able to synchronize exactly even in the presence of substitution errors.

\section{Proofs}  \label{sec:proofs}

\subsection{Proof of Theorem \ref{thm:indel}} \label{subsec:indel_proof}

We first prove part (b) of the theorem.

\emph{(b) (Expected communication required)}:
When there are $d$ deletions and $i$ insertions ($t=d+i$),  the {total} number of bits transmitted by the encoder to the decoder can be expressed as
\be \label{eq:nbits_12}
\begin{split}
  N_{1 \to 2} (d,i)& = N_a(d,i) +  N_h(d,i) +  N_v(d,i),
\end{split}
\ee
where $N_a, N_h$ and $N_v$ represent the number of bits sent for anchors, hashes, and VT syndromes, respectively. First, we will prove by induction that the expected total number of anchor bits can be bounded as 
\be \expec N_{a}(d,i) \leq  2(d+i-1)m_a. \label{eq:na_bound} \ee

The bound \eqref{eq:na_bound} holds for $(d=1, i=0)$
and $(d=0, i=1)$ since the encoder will start by sending the VT syndrome and a hash for $X$  if the length of $Y$ is $(n \pm 1)$. No anchor bits are required in this case. For $d+i>1$, we have the following contributions to $\expec N_a(d,i)$:
\begin{enumerate}
\item  If the length of $Y$ differs from $X$ by more than one, $m_a$ anchor bits are sent  in the first round.

\item When the decoder correctly matches the first set of anchor bits,  the  probability of $j$ deletions (out of $d$) and $k$ insertions (out of $i$) occurring to the left of the anchor is $\frac{1}{2^{d+i}}  {d \choose j} {i \choose k}$. This is because the locations of the edits are uniformly distributed, hence each edit is equally likely to be to the  left or to the right of the anchor. Therefore, when an unique match is found for the anchor bits, the  expected number of \emph{additional} anchor bits required in future rounds is 
\[    \sum_{j=0}^d \sum_{k=0}^i \frac{1}{2^{d+i}}  {d \choose j} {i \choose k} ( \expec N_a(j,k) + \expec N_a(d-j,i-k)).\] 

\item Anchor Edited: The decoder may fail to find a match for the set of anchor bits within the window of $\kappa \sqrt{n}$ bits due to one of the $m_a$ anchor bits undergoing an edit. 
Here $\kappa >0$ is a generic constant, whose exact value is not important. Since the probability of a given bit being edited is 
$\tfrac{t}{n}$, the probability of at least one of the anchor bits being edited is bounded by  $\frac{m_a t}{n}$. Since the decoder requests additional sets of anchor bits until it has identified a match, the expected  number of  additional anchor bits required in this case is bounded by 
\ben \begin{split} 
& \frac{m_a t}{n}  (m_a) + \left(\frac{m_a t}{n}\right)^2 (2m_a) +  \left(\frac{m_a t}{n}\right)^3 3 m_a \ldots \\
&  = m_a \frac{ m_a t/n}{ (1 - m_at/n)^2} <  m_a \frac{2m_a t}{n}, 
\end{split} \een
where the last inequality holds  because $\frac{t m_a}{n} \to 0$ as $n \to \infty$ as $t=o(n/\log n)$ and $m_a=c \log n$.

\item Unbalanced Edits: The decoder may fail to find a match for the set of anchor bits (within the window of $\kappa \sqrt{n}$ bits) if there are significantly more deletions/insertions on one side of the anchor  than the other. More precisely,  if  the number of deletions and insertions  to the left of the anchor in $X$ are denoted by $J$ and $K$, respectively,  the anchor in $Y$ will lie outside a window of $\kappa \sqrt{n}$ (centred at $(n-d+i)/2$) only if
\be \left| (J-K) - \frac{(d-i)}{2}\right| > \kappa \sqrt{n}. \label{eq:unbal_event} \ee
Since the locations of the edits are uniformly random, the probability of the event above can be bounded via a large deviations argument.
\begin{lem}
Let $J, K$ denote the number of deletions and insertions, respectively, to the left of the anchor in $X$. Then, for any $r >0$, the following holds for sufficiently large $n$:
\[ P\left( \left| (J-K) - \frac{(d-i)}{2}\right| > \kappa \sqrt{n} \right) \leq  n^{-r}.  \]
\label{lem:unbal_edits}
\end{lem}
\begin{IEEEproof} See Appendix \ref{app:unbal_edits_proof}. \end{IEEEproof}
Using the naive upper bound of $n$ for the extra bits required when the event in \eqref{eq:unbal_event} occurs,  Lemma \ref{lem:unbal_edits} implies that the expected number of extra bits required due to this event is bounded by $n^{-(r-1)}$ for sufficiently large $n$, where $r$ is a constant that can be chosen arbitrarily large.

\item An incorrect unique match for the anchor bits occurs when the anchor bits match with an independent substring of length $m_a$ within the window of 
$\kappa \sqrt{n}$ bits \emph{and} either one of the following occurs: a) there has been an edit in at least one of the true anchor bits, or b) the true set of anchor bits lies outside the window. 

Using the arguments in points 3) and 4) above for these events, the probability of an incorrect unique match is  bounded by 
\[ \frac{\kappa \sqrt{n}}{2^{m_a}}  \left(\frac{t m_a}{n}  + n^{-(r-1)} \right)= \kappa' \frac{ t m_a \sqrt{n}}{2^{m_a}} = \kappa' \frac{ t m_a}{n^{c+1/2}}  \]
for some $\kappa' >0$ since $r>0$ can be chosen to be a large positive constant. Bounding the extra bits required in the event of an incorrect unique match by  $n$, the expected number of additional bits due to this event is at most  \[ \kappa' \frac{ t m_a}{n^{c+1/2}} \cdot  n = \frac{\kappa' t m_a}{n^{c-1/2}} = o(1), \] since  $c > 1.5$ and $t=o(n/\log n)$.

\item Multiple matches for the anchor: The probability of having at least one independent substring of length $m_a$ within the window of $\kappa \sqrt{n}$ matching the anchor bits is  $\frac{\kappa \sqrt{n}}{2^{c \log n}}$. Bounding the number of additional bits required in this case by $n$, the expected number of extra bits due to multiple matches for the anchor is bounded by 
\[ \frac{\kappa \sqrt{n}}{2^{c \log n}} n =  \frac{\kappa}{n^{c-1.5}}  = o(1),  \]
since $c > 1.5$.
\end{enumerate}
Adding all the above contributions, the expected number of anchor bits required can be bounded as 
\be
\begin{split}
& \expec[ N_a(d,i)]    \leq \,  m_a  + \sum_{j=0}^d \sum_{k=0}^i \frac{1}{2^{d+i}}  {d \choose j} {i \choose k} ( \expec N_a(j,k) \\
&  \hspace{0.75in} + \expec N_a(d-j,i-k)) + \frac{2 t m_a}{n} m_a   + o(1).
\end{split}
\ee
Expanding the RHS, we obtain
\be
\begin{split}
& \expec[ N_a(d,i)]  \leq  \, m_a\left(1 + \frac{2 t m_a}{n}\right) + \frac{1}{2^{d+i}}\Bigg[ 2 \expec N_a(d,i)   \\ 
& +  \sum_{j=1}^{d-1} \sum_{k=1}^{i-1}  {d \choose j} {i \choose k} ( \expec N_a(j,k) + \expec N_c(d-j,i-k))  \\
& + \sum_{k=1}^i(\expec N_a(0,k) + \expec N_a(d,i-k)) {i \choose k} \\
& + \sum_{j=1}^d(\expec N_a(j,0) + \expec N_a(d-j,i)) {d \choose j}   \Bigg] + o(1). 
\end{split}
\label{eq:na_expand}
\ee

Assume towards induction that $ \expec N_a(j,k) < 2(j+k-1)m_a$ for all $j, k$ such that $j+k \leq (d+i-1)$. Using this in \eqref{eq:na_expand}, we obtain
\be
\begin{split}
&(1 - 2^{-(d+i-1)}) \expec N_a(d,i) \\
&  \leq  m_a \left( 1 + \tfrac{2 t m_a}{n}\right)  +  \frac{2(d+i-2)m_a}{2^{d+i}}\Bigg[ \sum_{k=1}^i {i \choose k}   \\
& \quad + \sum_{j=1}^d {d \choose j} + \sum_{j=1}^{d-1} \sum_{k=1}^{i-1} {d \choose j} {i \choose k}   \Bigg] + o(1) \\ 
& =   m_a \left( 1 + \tfrac{2 t m_a}{n}\right)  \\
& \  + \frac{2 (d+i-2)m_a}{2^{d+i}}(2^i  + 2^d -2 + (2^d-2)(2^i-2)) + o(1).
\end{split}
\label{eq:indel_recursion}
\ee
For $d+i> 1$, \eqref{eq:indel_recursion} implies  that
\be
\begin{split}
\expec N_a(d,i) &  < \frac{m_a \left( 1 + \tfrac{2 t m_a}{n}\right)}{1 - 2^{-(d+i-1)}} + 2 (d+i-2)m_a + o(1) \\
& <  2(d+i-1)m_a,
\end{split}
\ee
where the last inequality holds for large enough $n$ because $m_a = c \log n$ and $t=o(n/\log n)$, hence $\tfrac{2 t m_a}{n} \to 0$ as $n \to \infty$
This establishes \eqref{eq:na_bound}.

To upper bound the expected values of $N_h$ and $N_v$, we note that a hash is requested whenever the anchor bits indicate an offset of zero or one, and  a VT syndrome is requested whenever the anchor bits indicate an offset of one. Therefore the number of  times hashes (and VT syndromes) are requested by the decoder is bounded above by the number of times anchor bits are sent. Hence
\be
\expec N_h(d,i) < \expec \left[\tfrac{N_a(d,i)}{m_a}\right]{m_h} + m_h < 2(d+i-1)m_h +m_h.
\label{eq:nh_bound}
\ee
The additional  $m_h$ in the bound is to account for the fact that hashes and VT syndromes are sent in the beginning if the  length of $Y$ is either $n, n-1$, or $n+1$. Similarly,
\be \expec N_v(d,i) < \left(\expec\left[\tfrac{N_a(d,i)}{m_a}\right]  +1\right) \log n  < (2(d+i-1) +1)\log n. \label{eq:nv_bound}\ee
Combining \eqref{eq:na_bound}, \eqref{eq:nh_bound}, and \eqref{eq:nv_bound} and substituting $m_a=m_h=c \log n$ gives the upper bound on $\expec N_{1 \to 2}(d,i)$.

To bound $N_{2 \to 1}$, we first note that the information sent by the decoder back to the encoder consists of responses to anchor bits and hash bits. Each time the decoder receives a set of anchor bits, its response is either: a) Send additional anchor bits (i.e., no match found), or b) the instruction ``Split, x,y", where each of  x,y are the instructions for the pieces on either side of the anchor. Recall that $x,y$ can take one of three values: Verify, VT mode, or  Anchor. Thus  each time anchor-bits are sent, the decoder has to respond with one of $1 + 3 \times 3 =10$ possible options, which requires
four bits. Each time a hash is sent, the decoder needs to send back a one bit response (to indicate whether synchronized or not). Therefore the expected  number of bits sent by the decoder is
\be
\expec N_{2 \to 1} (d,i)  < 4 \cdot \frac{\expec N_a}{m_a} + 1\cdot \frac{\expec N_h}{m_h}  \leq 10(d+i-1) +1.
\ee
This completes the proof of part (b).

\emph{(a) (Probability of error)}:
An synchronization failure occurs if and only if two  non-identical substrings  are erroneously declared `synchronized' by a hash comparison. Denoting the  event of synchronization failure  by $\mc{E}$, we write
\be
P_e= P(\mc{E}) \leq P(\mc{E} |  \, \mc{F}) P(\mc{F}) + P(\mc{E} | \, \mc{F}^c),
\label{eq:Pe_exp}
\ee
where $\mc{F}$ denotes the event that at least one of the anchors was matched erroneously.

 First consider the second term $P(\mc{E}  | \, \mc{F}^c)$. As there are a total of $t$ edits and no anchor mismatches, in any step there can be at most $t$ substrings that are potential sources of error. Since any substring is sub-divided by an anchor at most $\log n$ times,  a union bound yields
\be
P(\mc{E} \mid \mc{F}^c) \leq {t \log n} \cdot P(\text{hash collision}) = \frac{t \log n}{n^c},
\label{eq:EFc}
\ee
where the last equality is due to the fact that a hash of length $c \log n$ drawn from a universal family of hash functions has collision probability $n^{-c}$ \cite{CarterWeg77}.

Next, we compute $P(\mc{F})$. The probability of an anchor mismatch in a piece of length $n/2^k$  is $\frac{\kappa \sqrt{n/2^k}}{n^c}$ because we search for a match within a window of size $\kappa \sqrt{n/2^k}$. Since there are at most $2^k$ unsynchronized pieces of length $n/2^k$, where $k=0,1,\ldots, (\log n) -1$, we have
\be 
P(\mc{F})  \leq \sum_{k=0}^{(\log n)-1} 2^k \ \frac{\kappa \sqrt{n/2^k}}{n^c} =  \frac{\kappa }{n^{c-\frac{1}{2}}}  \sum_{k=0}^{(\log n)-1} 2^{k/2} \leq \frac{3 \kappa}{n^{c-1}}.
\label{eq:PF}
\ee
Finally, $P(\mc{E} | \, \mc{F})$ is bounded as follows, noting that number of times anchor  bits are requested is at most $n/m_a$.
\be  \begin{split} P(\mc{E} | \, \mc{F}) = & (\text{number of times anchor bits are requested})  \\
& \times  P(\text{hash collision}) \\ 
&  \leq  \frac{n}{c \log n} \times \frac{1}{n^c}. \end{split} \label{eq:PE_F} \ee
Substituting \eqref{eq:EFc}, \eqref{eq:PF}, and \eqref{eq:PE_F} in \eqref{eq:Pe_exp} completes the proof.

\subsection{Proof of Theorem \ref{thm:sing_round}} \label{subsec:sing_round_proof}
($a$)  A piece remains unsynchronized at the end of the algorithm only if is there is a hash collision in one of the pieces, i.e., the hashes at the encoder and decoder agree despite their versions of the piece being different. With $c_h \log n $ hash bits, the probability of this event is $n^{-c_h}$ for each piece. Taking a union bound over the $n^{\bar{a}}$
pieces yields the result.

($b$)  In the first step, the number of bits sent by the encoder is deterministic: for each of the $n^{\bar{a}}$ pieces, it sends $m_a$ anchor bits, $m_h$ hash bits, and
$\log(n^a +1)$ bits for the VT syndrome.  The total number of bits sent by the encoder in the first  step  is  therefore
\be
N_{1 \to 2}^{(1)} = (c_a \log n + c_h \log n + \log (n^a +1) ) \, n^{\bar{a}}.
\label{eq:n12_first}
\ee
For each piece, the decoder sends  $1$ bit  back to the encoder to indicate whether the piece was synchronized or not. Thus the number of bits sent by the decoder is $n^{\bar{a}}$.

The number of bits sent by the encoder in  the final step is
\be
\begin{split}
N_{1 \to 2}^{(2)}  =  n^a & (\text{number of  pieces declared} \\
&  \text{unsynchronized after first step}).
\end{split}
\ee
A sufficient condition for a piece  to be declared synchronized after the first step is that it contains zero or one edits \emph{and}  the anchors on either side of the piece are aligned by the decoder in the correct position. (In addition, there may be some pieces declared synchronized due to a wrong anchor match followed by a hash collision, but we only want an upper bound for the number of bits sent in the final step.)

Since the locations of the edits are uniformly random, the probability of a piece containing none of the $n^b$ edits is
\be
p_0 = \left(1 - \frac{n^a}{n} \right)^{n^b},
\label{eq:p0}
\ee
and the probability of a piece undergoing exactly $1$ edit is
\be
p_1 = {n^b \choose 1} \frac{n^a}{n} \left(1 - \frac{n^a}{n} \right)^{n^b -1}.
\label{eq:p1}
\ee
  If the anchors for each of the $n^{\bar{a}}$ pieces were aligned by the decoder in the correct positions, then the expected number  of unsynchronized pieces after the first round would be $(1- p_0 - p_1)n^{\bar{a}}$.  We now argue that the  expected number of unsynchronized pieces after the first round is bounded by
\be \begin{split} 
&  \left[  1- p_0\left(1-  \frac{m_a n^b}{n}  - \frac{2n^{a/2}}{n^{c_a}} \right)  \right.\\
& \  \left. - p_1\left(1 - \frac{m_a}{n^a} - \frac{m_a n^b}{n}  - \frac{2n^{a/2}}{n^{c_a}} \right)  \right] n^{\bar{a}}. \end{split} 
  \label{eq:expec_unsync_pieces} \ee
A piece with zero edits remains unsynchronized after the first round only if one of the following occurs: a) the anchor at the end of the piece was either mismatched or failed to be matched due to an edit in the anchor, or  b) there were multiple matches for one of the anchors at either end of the piece. Since the locations of the $n^b$ edits are uniformly random, the probability of an anchor undergoing an edit is $m_a n^b/n$. The probability of multiple matches for an anchor of length $m_a=c_a \log n$ in a window of size $\sqrt{n^a}$ is $\sqrt{n^{a}}/{n^{c_a}}$. Using the union bound for the two anchors at either end of the piece yields a bound of $2 \sqrt{n^{a}}/{n^{c_a}}$ for the probability of the event b). Hence the probability of a piece with zero edits remaining unsynchronized after the first round is $(1-  \frac{m_a n^b}{n}  - \frac{2n^{a/2}}{n^{c_a}})$.

By a similar argument, the probability that a piece with one edit remains unsynchronized after the first round is  
$(1  - \frac{m_a n^b}{n}  - \frac{2n^{a/2}}{n^{c_a}} - \frac{m_a}{n^a})$, with the term $\frac{m_a}{n^a}$ being the  probability of the single edit being in one of the $m_a$ anchor positions. Thus \eqref{eq:expec_unsync_pieces} holds.

For sufficiently large $n$, the term $p_0 +p_1$ can be bounded from below as follows.
\be
\begin{split}
& p_0 + p_1   = \left(1 - \frac{n^a}{n} \right)^{n^b}  \left( 1 +  \frac{n^{b- \bar{a}}}{1- n^{-\bar{a}}}\right) \\
 & = \left( \left( 1 - n^{-\bar{a}}\right)^{n^{\bar{a}}} \right)^{n^{b-\bar{a}}}  \left( 1 +  \frac{n^{b- \bar{a}}}{1- n^{-\bar{a}}} \right) \\
& \stackrel{(a)}{>} \left( e^{-1}\left( 1 - \tfrac{1}{2} n^{-\bar{a}} - n^{-2\bar{a}}\right) \right)^{n^{b-\bar{a}}}  \left( 1 +  n^{b- \bar{a}} \right) \\
& \stackrel{(b)}{>}  \exp(- {n^{b-\bar{a}}}) (1- n^{b-2\bar{a}})(1 + n^{b-\bar{a}}).
\end{split}
\label{eq:p0p1bound}
\ee
In \eqref{eq:p0p1bound}, $(a)$ is obtained using the Taylor expansion of $(1+x)^{1/x}$ near $x=0$. $(b)$ holds because for large enough $n$
\be (1 - \tfrac{1}{2} n^{-\bar{a}} - n^{-2\bar{a}})  >  (1 -   \tfrac{2}{3} n^{-\bar{a}}) \ee
 and for $\bar{a} > b$
 \be (1 -   \tfrac{2}{3} n^{-\bar{a}})^{n^{b-\bar{a}}}  =  (1 -   \tfrac{2}{3} n^{-\bar{a}})^{1/n^{\bar{a}-b}} > \left(1 -   \frac{n^{-\bar{a}}}{n^{\bar{a}-b}} \right). \ee

Using the lower bound \eqref{eq:p0p1bound}  for $p_0+p_1$ in \eqref{eq:expec_unsync_pieces},  the expected number of bits sent by the encoder in the final step can be bounded as follows.
\be
\begin{split}
& \expec N_{1 \to 2}^{(2)}    \leq \left[1- p_0\left(1-  \frac{m_a n^b}{n}  - \frac{2n^{a/2}}{n^{c_a}} \right) \right.  \\
&  \hspace{0.75in} \left. - p_1\left(1 - \frac{m_a}{n^a} - \frac{m_a n^b}{n}  - \frac{2n^{a/2}}{n^{c_a}} \right)  \right] n^{\bar{a}} \cdot n^a \\
& \stackrel{(a)}{=} (1- p_0 - p_1)n + p_1 m_a n^{\bar{a}}  +  (p_0 +p_1) m_a n^b (1 + o(1))  \\
& \stackrel{(b)}{<}  \big[ 1 - \exp(- {n^{-(\bar{a}-b)}}) (1 + n^{-(\bar{a}-b)})(1- n^{-(2\bar{a}-b)}) \big]n \\
& \quad + m_a n^b + m_a n^b (1 + o(1)) \\
& \stackrel{(c)}{<}  \big[ 1 - \big(1- n^{-(\bar{a}-b)} + \tfrac{1}{2} n^{-2(\bar{a}-b)}  - \tfrac{1}{6} n^{-3(\bar{a}-b)}\big) \\
& \qquad  \cdot   (1 + n^{-(\bar{a}-b)})(1- n^{-(2\bar{a}-b)}) \big] n + 2 m_a n^b (1 + o(1)) \\
& = \tfrac{1}{2} n^{1-2(\bar{a}-b)} + n^{ a- (\bar{a}-b) } + O( n^{1-3(\bar{a}-b)} ) \\
& \ + 2 m_a n^b (1 + o(1)) 
 = \left( \tfrac{1}{2} n^{1-2(\bar{a}-b)}  + 2 m_a n^b \right)(1 + o(1)).
\end{split}
\label{eq:n12_second}
\ee
In the chain above, $(a)$ holds because  $c_a > 1 + \tfrac{a}{2}$; $(b)$ is obtained by using  the lower bound \eqref{eq:p0p1bound} for $p_0 + p_1$, and the upper bounds $p_1< n^a n^b /n$ and $(p_0 +p_1) <1$; for $(c)$ we have used the inequality \[ e^{-x}  > 1 - x + \frac{x^2}{2} -\frac{x^3}{6} \ \text{ for } x>0,\]
and the fact that $p_1 < n^b /n$. Combining \eqref{eq:n12_second} with \eqref{eq:n12_first} completes the proof.

\subsection{Proof of Theorem \ref{thm:single_burst}}  \label{subsec:sing_burst_proof}
In the first step, the encoder sends the VT syndrome of substrings $X^1$ and $X^B$, which require $\log(1 +n/B)$ bits each. Thus $ N_{1 \to 2}$ equals the sum of  $2\log (1 +n/B)$ and the bits transmitted by the encoder in the second step. Recall that the latter is given by $(j^*- l^* +1)(B-2)$, with $j^*, l^*$ defined in \eqref{eq:jsls}.

The lower bound is obtained by considering the ideal case where the single edits in both $X^1$ and $X^B$ occur in runs of length one, i.e., $j_1=l_1$, and $j_B=l_B$. In this case, there are two possibilities:

1) The starting position of the burst edit in $X$ is of the form $aB +1$ for some integer $a \geq 0$,  in which case the  edit will be in the $(a+1)$th bit of \emph{all} substrings $X^k$, $1 \leq k \leq B$. The encoder then only needs to send $1$ bit/substring in the final step.

 2) The starting position of the burst edit in $X$ is of the form $aB +q$ for $2\leq q \leq B$, then $j_B=j_1-1$, i.e.,  the position of the edit in $X^B$ is one less than the position in $X^1$. Here two bits/substring are needed in the final step.
 
As the starting position of the burst is uniformly random, the average number of bits per substring in the ideal case is
\be \frac{1}{B} \cdot 1 + \left(1-\frac{1}{B}\right) \cdot 2 = 2- \frac{1}{B} \ee
Hence the expected number of bits sent in the final step for substrings $X^2, \ldots,X^{B-1}$ is lower bounded by $(2-\tfrac{1}{B})(B-2)$.

To obtain an upper bound on $(j^*- l^* +1)$, we start by observing that
\be (l^* - j^*) \leq l_1 -j_1 +1, \ \  (l^* - j^*) \leq l_B -j_B +1,  \label{eq:ubjl}\ee
which follows directly from \eqref{eq:jsls}.  Note that $(l_1 -j_1 +1)$  and $(l_B -j_B +1)$ are the lengths of the runs containing the edit in $X^1$ and $X^B$, respectively.  Denoting these by $R_1$ and $R_B$, \eqref{eq:ubjl} can be written as
\be  (l^* - j^*) \leq \min\{R_1, R_B\}.  \label{eq:ubjl1} \ee
Since the binary string $X$ is assumed to be uniformly random, the bits in each substring are i.i.d Bernoulli$(\tfrac{1}{2})$.  $R_1$ and $R_B$ are i.i.d, and their distribution is that of a run-length \emph{given} that one of the bits in the run was deleted. This distribution is related to the inspection paradox and it can be shown \cite{RossInsp} that as $n$ grows large, the probability mass function converges to
\be
P(R_1 = r) = P(R_B = r) = r \cdot  2^{-(r+1)}, \quad r=1,2,\ldots
\ee
Therefore, for $r \geq 1$,
\be
\begin{split}
 P(\min\{R_1, R_B\}  \geq r)  & = P(R_1 \geq r) \cdot P(R_B \geq r) \\
& = (2^{-r}(1+r))^2(1+o(1)).
\end{split}
\label{eq:minr_dist}
\ee
The expected number of bits required per substring in the final step can be bounded by using \eqref{eq:minr_dist} in \eqref{eq:ubjl1}: 
\be
\begin{split}
&\expec [ l^* - j^* +1] \leq \expec[ \min\{R_1, R_B\} ] +1 \\
& = 1+ \sum_{r\geq 1}4^{-r}(1+r)^2 (1+o(1)) =1+ \tfrac{53}{27}(1+o(1)).
\end{split}
\label{eq:runmin_dist}
\ee
Thus for sufficiently large $n$, $\expec N_{1 \to 2}$ can be bounded as
\be
\begin{split}
\expec N_{1 \to 2} & = 2\log (1 +n/B)  + \expec [ l^* - j^* +1] (B-2)\\
&  \leq 2\log (1 +n/B)  + 3(B-2).
\end{split}
\ee

To compute $\expec[N_{2 \to 1}] $, recall that the decoder sends the index of the run containing the edit in the first and last substrings. Each of these substrings is a binary string of length $n/B$ drawn uniformly at random. Hence the expected number of runs in each substring is $n/(2B)$, and the expected number of bits required to indicate the index of a run in each string is $\log(\tfrac{n}{2B})$.

\section{Discussion} \label{sec:concl}
The interactive algorithm (and its single-round adaptation) can be extended to synchronize strings over non-binary discrete alphabets---strings of ASCII characters, for example---that differ by $o(\tfrac{n}{\log n})$ indel edits.  This is done by replacing the binary VT code with a $q$-ary VT code \cite{Tenengolts84}, where $q$ is the alphabet size.  The performance of the synchronization algorithm for a $q$-ary alphabet is discussed in \cite{DolBit13}, and the simulation results reported in \cite{BitouzeSYD13} demonstrate significant communication savings over rsync.

We now discuss some directions for future work. The multi-round algorithm of Section \ref{sec:insdel} and the one-round algorithm of Section \ref{sec:limited_rounds} represent two extreme points of the trade-off between the number of rounds and the total communication required for synchronization. In general, one could have an algorithm which takes up to $r$ rounds, where $r$ is a user-specified number. Designing such an algorithm, and determining the trade-off between $r$ and the total communication required is an interesting open question.

The simulation results in Section \ref{sec:bursts} show that a guess-and-check approach is effective when there are multiple bursty indel edits. An important open problem is to obtain theoretical bounds on the expected communication of the algorithm when there are multiple bursts of varying length. Investigating the performance of the synchronization algorithm for non-binary strings with bursty edits is another problem of practical significance.

When the edits are a combination of indels and substitutions, Table \ref{table:sub_edits} shows that  synchronizing to within a small Hamming distance requires very little communication as long as the number of indel edits is small. A complete system for perfect synchronization could first invoke the synchronization algorithm with a distance estimator hash, and then use LDPC syndromes as an ``outer code" to achieve perfect synchronization. If the normalized Hamming distance at the end of the first step is $p$, an ideal syndrome-based algorithm would need $nH_2(p)$ bits in the second step to achieve exact synchronization. ($H_2$ is the binary entropy function.) For the example in Table \ref{table:sub_edits} with $n=10^6$ and $40$ hash bits, the final normalized distance $p$  is less than $3.5 \times 10^{-4}$, which implies that fewer than $0.46 \%$ additional bits are required for perfect synchronization. Building such a complete synchronization system, and integrating the techniques presented here into practical applications such as video synchronization is part of future work.

\appendix
\subsection{Proof of Lemma \ref{lem:run_lem}} \label{app:run_lem_proof}
Consider a length $m$ binary sequence $Z=(Z_1, \ldots, Z_m)$ where the $Z_i$ are i.i.d. Bernoulli$(1/2)$ bits. For $i=2, \ldots,m$ define  random  variable $U_i$ as follows: $U_i=1$ if $Z_i \neq Z_{i-1}$ and $U_i=0$ otherwise. Then the number of runs in $Z$ can be expressed as $1 + U_2 \, + \, U_3 \,+ \ldots + \, U_m$. 
Note that $U_i$ are i.i.d. Bernoulli$(1/2)$ bits due to the assumption on the distribution of $Z$. Hence
\be
\begin{split}
& P(Z \text{ has fewer than } \tfrac{m}{2}(1-\delta)  \text{ runs}) \\
& = P(U_2 + \ldots + U_m  < \tfrac{m}{2} (1 - \delta) - 1)  \\
& \stackrel{(a)}{\leq}  e^{-(m \delta +1)^2/2(m-1)} < e^{-(m-1) \delta^2/2}.
\end{split}
\label{eq:run_chernoff}
\ee
In \eqref{eq:run_chernoff}, ($a$) is obtained using Hoeffding's inequality \cite{mitzbook} for i.i.d. Bernoulli random variables.

Now set $e^{-(m-1)\delta^2/2} = \e$ so that $\delta = \sqrt{\frac{2}{m-1} \ln \frac{1}{\e}}$. Let $\mc{A}_\delta$ be the set of length $m$ binary sequences with at least $\frac{m}{2}(1-\delta)$ runs, and let $\mc{A}^c_\delta$ denote its complement. Then from \eqref{eq:run_chernoff} we have
\be
\begin{split}
\e = e^{-(m-1)\delta^2/2} \, > \, \sum_{z \in \mc{A}^c_\delta} P(z) = \sum_{z \in \mc{A}^c_\delta} \, 2^{-m} = \abs{\mc{A}^c_\delta} 2^{-m}.
\end{split}
\ee
It follows that $\abs{\mc{A}^c_\delta} < 2^m \e$, or  $\abs{\mc{A}_\delta} \geq 2^{m}(1-\e)$. Thus we have constructed a set $\mc{A}_\delta$ with at least $2^{m}(1-\e)$ sequences, each having at least $\frac{m}{2}(1-\delta)$ runs.

\subsection{Proof of Proposition \ref{prop:sketch_chernoff}} \label{app:sketch_chernoff_proof}

 The estimator $\hat{d}_H(x,y)$ is a strictly increasing function of $\bar{Z}$. Therefore the event $\{ \bar{Z} >  p + \delta \}$ is equivalent to 
 \be \left\{  \hat{d}_H(x,y)   > \frac{ \ln(1 - 2(p+\delta))}{\ln(1 - \kappa/n)} \right\}.  \label{eq:event1} \ee
 Using \eqref{eq:Zi_dist} to write $d_H(x,y) = \tfrac{\ln(1 - 2p)}{\ln(1 - \kappa/{n})}$, we have 
 \be
 \begin{split}
& \frac{ \ln(1 - 2(p+\delta))}{\ln(1 - \kappa/n)}  =  d_H(x,y) + \frac{\ln(1 - 2 {\delta}/(1-2p))}{\ln(1 - \kappa/n)} \\
&  \geq  d_H(x,y) + \frac{2 \delta }{(1-2p)} \frac{n}{\kappa}  + \frac{2 \delta^2 }{(1-2p)^2} \frac{n}{\kappa},
 \label{eq:event1_eq}
 \end{split}
 \ee
where the inequality is obtained using Taylor's theorem for $\ln(1- x)$. 

Similarly, the event $\{ \bar{Z} <  p - \delta \}$ is equivalent to
 \be \left\{  \hat{d}_H(x,y)   <  \frac{ \ln(1 - 2(p-\delta))}{\ln(1 - \kappa/n)} \right\}.  \label{eq:event2} \ee
As before, the RHS can be written as
\be
\begin{split}
& \frac{ \ln(1 - 2(p-\delta))}{\ln(1 - \kappa/n)}  = d_H(x,y) + \frac{\ln(1 + 2 \delta/(1-2p))}{\ln(1 - \kappa/n)} \\
& \leq d_H(x,y) -\frac{2 \delta}{(1-2p)}\frac{n}{\kappa}  +  \frac{2 \delta^2 }{(1-2p)^2} \frac{n}{\kappa}.
\end{split}
 \label{eq:event2_eq}
\ee
Noting that $\bar{Z}$ is the empirical average of $m_h$ i.i.d Bernoulli random variables with mean $p$, the result is obtained by using Hoeffding's inequality \cite{mitzbook} to bound the probability of the events $\{ \bar{Z} >  p + \delta \}$ and $\{ \bar{Z} <  p - \delta \}$, which are equivalent to the events in \eqref{eq:event1} and \eqref{eq:event2}, respectively.

\subsection{Proof of Lemma \ref{lem:unbal_edits}} \label{app:unbal_edits_proof}

Using the triangle inequality, we have
\be
\begin{split}
 P\left( \left| (J-K) - \frac{(d-i)}{2}\right| > \kappa \sqrt{n} \right) &  \leq P\left( \left| J - \frac{d}{2}\right| > \frac{\kappa \sqrt{n}}{2} \right) \\
+ &  P\left( \left| K - \frac{i}{2}\right| > \frac{\kappa \sqrt{n}}{2} \right).
\end{split}
\label{eq:triangle_bound}
\ee
The random variable $J$, which is the number of deletions in the left half of $X$, can be expressed as a sum of $d$ indicator random variables as 
\be
J= U_1 + \ldots + U_d,
\ee 
where for $1 \leq  \ell \leq d$, $U_\ell=1$ if the $\ell$th deletion occurred in the left half of $X$, and $U_\ell = 0$ otherwise. Since the locations of the deletions are uniformly random, $P(U_\ell = 1) = P(U_\ell =0) = \frac{1}{2}$. Using Hoeffding's inequality \cite{mitzbook}, we have for any $\e >0$,
\be
\begin{split}
 P\left( \left| J - \frac{d}{2}\right| > d \e \right) & =  P\left( \left| \sum_{\ell =1}^d U_\ell - \frac{d}{2}\right| > d \e \right) \\
 & < 2 \exp(-2d\e^2).
 \end{split}
\ee
Substituting $\e = \frac{\kappa \sqrt{n}}{2d}$, we obtain
\be
 P\left( \left| J - \frac{d}{2}\right| > \frac{\kappa \sqrt{n}}{2}  \right)  < 2 \exp\left(-\frac{\kappa^2 n}{2d}\right).
 \label{eq:Jbound}
\ee
Using a similar argument for insertions, it follows that 
\be
 P\left( \left| I - \frac{i}{2}\right| > \frac{\kappa \sqrt{n}}{2}  \right)  < 2 \exp\left(-\frac{\kappa^2 n}{2i}\right).
 \label{eq:Kbound}
\ee
Since $t = d+i = o(n/\log n)$, for sufficiently large $n$ the exponents in \eqref{eq:Jbound} and \eqref{eq:Kbound}  satisfy
\be
\frac{\kappa^2 n}{2d} > r_0 \log n, \quad \frac{\kappa^2 n}{2i} > r_0 \log n,
\ee
for any constant $r_0 >0$. For any $r >0$, we can choose $r_0$ large enough so that the RHS of \eqref{eq:Jbound} and \eqref{eq:Kbound} are each less than 
$\tfrac{1}{2}n^{-r}$. Using these bounds in \eqref{eq:triangle_bound} completes the proof.

\section*{Acknowledgement}
The authors would like to thank H. Zhang for simulations of early versions of the interactive synchronization algorithm. They also  thank the anonymous reviewers whose comments led to a much improved paper.

\IEEEtriggeratref{8}


\begin{thebibliography}{24}
\bibitem{rsync}
A.~Tridgell and P.~Mackerras, ``The rsync algorithm.'' http://rsync.samba.org/,
  Nov 1998.

\bibitem{evf98}
A.~V. Evfimievski, ``A probabilistic algorithm for updating files over a
  communication link,'' in {\em Proc. ACM-SIAM
  Symposium on Discrete Algorithms}, pp.~300--305,  1998.

\bibitem{CormodePSV00}
G.~Cormode, M.~Paterson, S.~C. Sahinalp, and U.~Vishkin, ``Communication
  complexity of document exchange,'' in {\em Proc. ACM-SIAM Symp. on Discrete
  Algorithms}, pp.~197--206, 2000.

\bibitem{OrlitskyV01}
A.~Orlitsky and K.~Viswanathan, ``Practical protocols for interactive
  communication,'' in {\em Proc. IEEE Int. Symp. Information Theory},
  pp.~24--29, June 2001.

\bibitem{Trachten06}
S.~Agarwal, V.~Chauhan, and A.~Trachtenberg, ``Bandwidth efficient string
  reconciliation using puzzles,'' {\em IEEE Trans. Parallel Distrib. Syst.},
  vol.~17, no.~11, pp.~1217--1225, 2006.

\bibitem{DolBit13}
N.~Bitouz{\'e} and L.~Dolecek, ``Synchronization from insertions and deletions
  under a non-binary, non-uniform source,'' in {\em Proc. IEEE Int. Symp. Inf.
  Theory}, 2013.

\bibitem{MaKTse12}
N.~Ma, K.~Ramchandran, and D.~Tse, ``A compression algorithm using mis-aligned
  side-information,'' in {\em Proc. IEEE Int. Symp. Inf. Theory}, 2012.

\bibitem{VT65}
R.~R. Varshamov and G.~M. Tenengolts, ``Codes which correct single asymmetric
  errors,'' {\em Automatica i Telemekhanica}, vol.~26, no.~2, pp.~288--292,
  1965.
\newblock (in {R}ussian), English Translation in \emph{Automation and Remote
  Control}, ({\bf{26}}, No. 2, 1965), 286-290.

\bibitem{ZhangYR12}
H.~Zhang, C.~Yeo, and K.~Ramchandran, ``Vsync: Bandwidth-efficient and
  distortion-tolerant video file synchronization,'' {\em IEEE Trans. Circuits
  Syst. Video Techn.}, vol.~22, no.~1, pp.~67--76, 2012.

\bibitem{YazdiDol14}
S.~M.~S. {Tabatabaei Yazdi} and L.~Dolecek, ``A deterministic polynomial-time
  protocol for synchronization from deletions,'' {\em IEEE Trans. Inf. Theory},
  vol.~60, pp.~397--409, Jan. 2014.

\bibitem{sumilenkovic14}
L.~Su and O.~Milenkovic, ``Synchronizing rankings via interactive
  communication,'' in {\em Proc. IEEE Int. Symp. Inf. Theory}, 2014.

\bibitem{Wyner74}
A.~Wyner, ``Recent results in the {S}hannon {T}heory,'' {\em IEEE Trans. Inf. Theory}, vol.~20, no.~1, pp.~2--10, 1974.

\bibitem{Orlitsky93}
A.~Orlitsky, ``Interactive communication of balanced distributions and of
  correlated files,'' {\em SIAM J. Discrete Math.}, vol.~6, no.~4,
  pp.~548--564, 1993.

\bibitem{PradhanR03}
S.~S. Pradhan and K.~Ramchandran, ``Distributed source coding using syndromes
  ({DISCUS}): design and construction,'' {\em IEEE Trans. Inf. Theory},
  vol.~49, no.~3, pp.~626--643, 2003.

\bibitem{XiongDSC04}
Z.~Xiong, A.~Liveris, and S.~Cheng, ``Distributed source coding for sensor
  networks,'' {\em IEEE Sig. Proc. Magazine}, vol.~21, pp.~80--94, Sept.
  2004.

\bibitem{KTracht13}
A.~Kontorovich and A.~Trachtenberg, ``String reconciliation with unknown edit
  distance,'' in {\em Proc. IEEE Int. Symp. on Inf. Theory}, pp.~2751--2755,
  2012.

\bibitem{BitouzeSYD13}
N.~Bitouz{\'e}, F.~Sala, S.~M. S.~T. Yazdi, and L.~Dolecek, ``A practical
  framework for efficient file synchronization,'' in {\em Proc. 51st Annual
  Allerton Conf. on Comm., Control, and Computing}, pp.~1213--1220, 2013.

\bibitem{Lev65}
V.~I. Levenshtein, ``Binary codes capable of correcting deletions, insertions
  and reversals,'' {\em Doklady Akademii Nauk SSSR}, vol.~163, no.~4,
  pp.~845--848, 1965.
\newblock (in {R}ussian), English Translation in \emph{Soviet Physics Dokl.},
  (No. 8, 1966), 707-710.

\bibitem{Sloane00}
N.~J.~A. Sloane, ``On single-deletion-correcting codes,'' in {\em Codes and
  Designs, Ohio State University (Ray-Chaudhuri Festschrift)}, pp.~273--291,
  2000.
\newblock Online: http://www.research.att.com/~njas/doc/dijen.ps.

\bibitem{CarterWeg77}
J.~L. Carter and M.~N. Wegman, ``Universal classes of hash functions,'' {\em
  Journal of Comp. and Sys. Sci.}, vol.~18, pp.~143--154, April 1979.

\bibitem{Kushil98}
E.~Kushilevitz, R.~Ostrovsky, and Y.~Rabani, ``Efficient search for approximate
  nearest neighbor in high dimensional spaces,'' {\em SIAM Journal on
  Computing}, pp.~457--474, 2000.

\bibitem{RossInsp}
S.~M. Ross, ``The inspection paradox,'' {\em Probab. Eng. Inf. Sci.}, vol.~17,
  pp.~47--51, Jan. 2003.

\bibitem{Tenengolts84}
G.~Tenengolts, ``Nonbinary codes, correcting single deletion or insertion,''
  {\em IEEE Trans on Inf. Theory}, vol.~30, no.~5, pp.~766--, 1984.

\bibitem{mitzbook}
M.~Mitzenmacher and E.~Upfal, {\em Probability and Computing: Randomized
  algorithms and probabilistic analysis}.
\newblock Cambridge Univ. Press, 2005.

\end{thebibliography}
\end{document}